\documentstyle[aps,prl,preprint,floats,epsfig]{revtex}

\def\etal{{\it et al.}}

\begin{document}

\preprint{\tighten\vbox{\hbox{\hfil CLNS 01/1724}
                        \hbox{\hfil CLEO 01-5}}}

\title{
First Observation of $\overline{B}\to D^{(*)}\rho'^-$, $\rho'^-\to\omega\pi^-$}
\author{CLEO Collaboration}
\date{March 9, 2001}

\maketitle
\tighten

\begin{abstract} 
We report on the observation of  $\overline{B}\to D^{*}\pi^+\pi^-\pi^-
\pi^0$ decays. 
The branching ratios for $D^{*+}$ and $D^{*0}$ are  
(1.72$\pm$0.14$\pm$0.24)\%
and (1.80$\pm$0.24$\pm$0.27)\%,
respectively. Each final state has a 
$D^{*}\omega\pi^-$ component, with branching ratios 
(0.29$\pm$0.03$\pm$0.04)\% and (0.45$\pm$0.10$\pm$0.07)\% for the 
$D^{*+}$ and $D^{*0}$ modes, respectively.  We also observe 
$\overline{B}\to D\omega\pi^-$ decays.   
The branching ratios for $D^{+}$ and $D^0$ are  
(0.28$\pm$0.05$\pm$0.04)\% and (0.41$\pm$0.07$\pm$0.06)\%,  
respectively. A spin parity analysis of the $\omega\pi^-$ system
in the $D\omega\pi^-$ final
state shows a preference for a wide $1^-$ resonance. A fit to the $\omega\pi^-$ mass
spectrum finds a central mass of  (1349$\pm$25$^{+10}_{-5}$) MeV and width of 
(547$\pm$86$^{+46}_{-45}$) MeV.
We identify this object as the $\rho$(1450) or the $\rho'$.   
\end{abstract}
\newpage

{
\renewcommand{\thefootnote}{\fnsymbol{footnote}}


\begin{center}
J.~P.~Alexander,$^{1}$ C.~Bebek,$^{1}$ B.~E.~Berger,$^{1}$
K.~Berkelman,$^{1}$ F.~Blanc,$^{1}$ V.~Boisvert,$^{1}$
D.~G.~Cassel,$^{1}$ P.~S.~Drell,$^{1}$ J.~E.~Duboscq,$^{1}$
K.~M.~Ecklund,$^{1}$
% ...
\end{center}
 
\small
\begin{center}
$^{1}${Cornell University, Ithaca, New York 14853}\\
% ...
$^{5}${Carleton University, Ottawa, Ontario, Canada K1S 5B6 \\
and the Institute of Particle Physics, Canada}\\
% ...
$^{10}${Ohio State University, Columbus, Ohio 43210}\\
% ...
\end{center}
 
\setcounter{footnote}{0}
}
\newpage

\section{Introduction}\label{sec:Introduction}

Currently most $B$ meson decays are unknown.
The decay width is comprised of hadronic decays, leptonic decays and semileptonic decays. Leptonic decays are predicted to be very small, $\approx 10^{-4}$ in branching fraction.
The semileptonic branching ratio for $\overline{B}\to Xe^-\nu$, $X\mu^-\nu$,
and $X\tau^-\nu$ totals approximately
25\% \cite{PDG}. The remainder must come from hadronic decays, where the measured exclusive branching ratios total only a small fraction of the hadronic width.

To be more explicit, the measured hadronic decay modes for the 
$\overline{B}^0$
including $D^+ (n\pi^-)$, $D^{*+} (n\pi^-)$, where $3\ge n \ge 1$, 
$D^{+(*)}D_s^{-(*)}$, and $J/\psi$ exclusive totals only about 
10\% \cite{PDG}. The $B^-$ modes total about 12\%.

Yet, understanding hadronic decays of the $B$ is crucial to ensuring 
that decay modes used for measurement of CP violation truly 
exhibit the underlying quark decay mechanisms expected 
theoretically.

It is also interesting to note that
the average charged multiplicity in a hadronic $B^0$ decay is 
5.8$\pm 0.1$ 
\cite{chargedmult}. Since this multiplicity contains 
contributions from the decay of
$D^+$ or $D^{*+}$ normally present in $\overline{B}^0$ decay, we 
expect a sizeable, approximately several percent, 
decay rate into final states with four pions \cite{Argus}.
The seen $D^{(*)} (n\pi)^-$ final states for $n \leq 3$ are 
consistent with 
being quasi-two-body final states. For $n$ of two the $\rho^-$ 
dominates, while
for $n$ of three the $a_1^-$ dominates \cite{BigB}. These decays 
appear to occur from a simple spectator mechanism where the 
virtual $W^-$ materializes as a single hadron: $\pi^-$, $\rho^-$ 
or $a_1^-$. The decay rates can be understood in a simple ``factorization" 
model where the decay rate is given by the product between two currents, one between the $\bar{B}$ and the $D$ and
the other given by the virtual $W^-$ transforming into the light hadron of interest \cite{factorization}. 

In this paper we investigate final states for $n$ of 
4. We will show a large signal for the $D^{*+}\pi^+\pi^-\pi^-\pi^0$ 
final state in Section~\ref{sec:Dstarp4pi}. In Section~\ref{sec:Dspomegapi}
we will show that a substantial fraction, $\sim$20\%,
arise from $D^{*+}\omega\pi^-$ decays and that the $\omega\pi^-$ mass
distribution has a resonant structure around 1.4 GeV with a width
of about 0.5 GeV. In Section~\ref{sec:Ds04pi}, similar conclusions are
drawn about the $D^{*0}\pi^+\pi^-\pi^-\pi^0$ final state.
 The same structure is shown to exist  in
$D\omega\pi^-$ final states (Section~\ref{sec:Domegapi}) and we will use these
events to show in Section~\ref{sec:Dangular} that
the spin-parity is most likely $1^-$. This state is identified as the 
$\rho'$, sometimes called the $\rho$(1450). We perform a
generalized Breit-Wigner fit to get the best values of the mass and
width, discussed in Section~\ref{sec:mass} and the Appendix.

Other resonant substructure is searched for, but not found
(Section~\ref{sec:nullsearch}). Finally we summarize our findings and compare
with the predictions of factorization and other models in Section~\ref{sec:conclusions}. 

The data sample consists of 9.0 fb$^{-1}$ 
of integrated luminosity taken with the CLEO II and II.V
detectors \cite{CLEOdetector} using the CESR $e^+e^-$ storage ring 
on the peak of the 
$\Upsilon(4S)$ resonance and 4.4 fb$^{-1}$ in the
continuum at 60 MeV less center-of-mass energy. The sample 
contains 19.4 million $B$ mesons.

\section{Common Selection Criteria}\label{sec:Common}

Hadronic events are selected by requiring a minimum of five charged tracks, total
visible energy greater than 15\% of the center-of-mass energy, and a charged
track vertex consistent with the nominal interaction point.
To reject continuum we require that the Fox-Wolfram moment $R_2$ 
be less than 0.3 \cite{Fox-Wolf}. 

Track candidates are required to pass through a common spatial 
point defined by the origin of all tracks. Tracks with momentum below 
900 MeV/c are required to have an ionization loss in the drift 
chamber within $3\sigma$ of that expected for their assigned mass.\footnote{Here and throughout this paper $\sigma$ indicates an r.m.s. error.} (These 
requirements are not imposed on slow charged pions from $D^{*+}$ 
decay.) Photon candidates are required
to be in the ``good barrel region," within 45$^{\circ}$ of 
the plane perpendicular to the beam line that passes through the interaction
point, and have an energy distribution in 
the CsI calorimeter consistent with that of an electromagnetic 
shower. To select $\pi^0$'s,
we require that the diphoton invariant mass be between  -3.0 to 
+2.5$\sigma$ of the $\pi^0$ mass, where $\sigma$ varies with momentum and has an
average value of approximately 5.5 MeV. The two-photon candidates
are then kinematically fit by constraining their invariant mass be equal to the nominal $\pi^0$ mass.

We select $D^0$ and $D^+$ candidates via the decay modes shown in 
Table~\ref{table:massres}. 
We require that the invariant mass of the $D$ candidates lie 
within $\pm 2.5\sigma$ of the known $D$ masses. The $\sigma$'s are 
also listed in Table~\ref{table:massres}. The $D^0$ widths vary with 
the $D^0$ momentum, $p$, (units of MeV), while the $D^+$ widths are
not momentum dependent.   

We select $D^{*+}$ candidates by imposing the addition requirement
that the mass difference between $\pi^+ D^0$ and 
$D^0$ combinations is within $\pm 2.5\sigma$ of the known mass 
difference.
For the $D^{*0}$, we use the same requirement for the $\pi^0 D^0$ decay. The mass difference resolutions are 
0.63 and 0.90 MeV, for the $\pi^+D^0$ and $\pi^0 D^0$ modes, 
respectively \cite{Differences}.

\begin{table}[hbt]
\begin{center}
\caption{Mass Resolutions ($\sigma$) in MeV  ($p$ in units of MeV)}
\label{table:massres}
\begin{tabular}{cccc}
$D^+\to K^-\pi^+\pi^+$ &  $D^0\to K^-\pi^+$ & 
$D^0\to K^-\pi^+\pi^0$ & $D^0\to K^-\pi^+\pi^+\pi^-$ \\
6.0 & $p\times$0.93$\times 10^{-3}$+6.0 & 
$p\times$0.68$\times 10^{-3}$+11.6 & 
$p\times$0.92$\times 10^{-3}$+4.7 \\
\end{tabular}
\end{center}
\end{table}

 \section{Observation of $\overline{B}^0\to D^{*+}\pi^+\pi^-\pi^-
\pi^0$ Decays}\label{sec:Dstarp4pi}

\subsection {$\overline{B}$ Candidate Selection}

We start by investigating the $D^{*+}(4\pi)^-$ final 
state.\footnote{In this paper $(4\pi)^-$ will always denote the 
specific combination $\pi^+\pi^-\pi^-\pi^0$.} 
The $D^{*+}$  candidates are pooled with all combinations of
$\pi^+\pi^-\pi^-\pi^0$ mesons. 

Next, we
calculate the difference between the beam energy, $E_{beam}$, and 
the measured energy
of the five particles, $\Delta E$. The ``beam constrained" 
invariant mass of the $\overline{B}$
candidates, $M_B$, is computed from the formula
\begin{equation}
M_B^2=E_{beam}^2-(\sum_i\overrightarrow{p_i})^2~~~.
\end{equation}

To further reduce backgrounds we define
\begin{equation}\label{eq:chisq}
\chi_b^2=\left({{\Delta M_{D^*}}\over {\sigma(\Delta 
M_{D^*})}}\right)^2 + 
\left({{\Delta M_{D}}\over {\sigma(\Delta M_{D})}}\right)^2 + 
\sum_{n(\pi^0)}\left({{\Delta M_{\pi^0}}\over {\sigma(\Delta 
M_{\pi^0})}}\right)^2~~~,
\end{equation}
where $\Delta M_{D^*}$ is the computed $D^*-D^0$ mass difference
minus the nominal value, $\Delta M_{D}$ is
the invariant candidate $D^0$ mass minus the known $D^0$ mass and 
$\Delta M_{\pi^0}$ is the measured $\gamma\gamma$ invariant mass 
minus the
known $\pi^0$ mass. All $\pi^0$'s in the final state are included 
in the sum.
The $\sigma$'s are the measurement errors. We select candidate 
events in each mode 
requiring that  $\chi^2_b < C_n$, where $C_n$
varies for each decay $D^0$ decay mode. For the $Kn\pi$ decay 
modes we use
$C_n=~12,$ 8, and 6, respectively.

\begin{figure}[bht]
\centerline{\epsfig{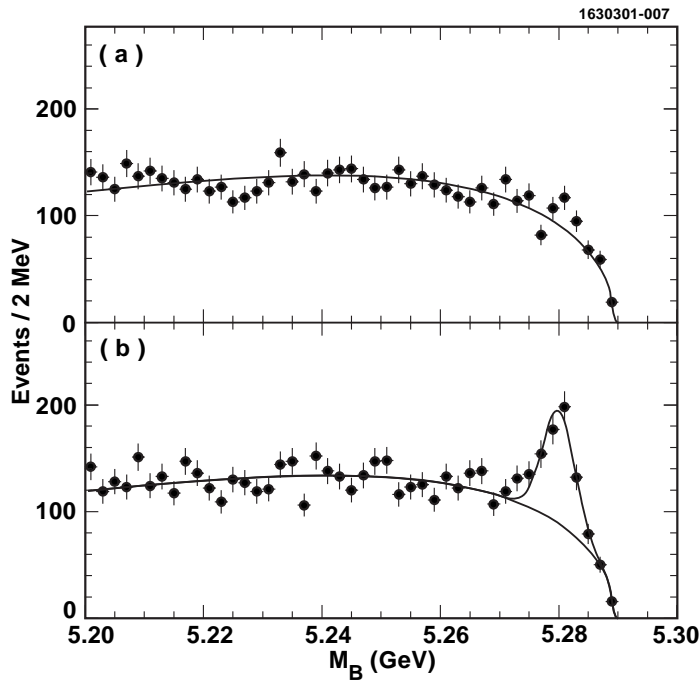}}
\caption{\label{bm_4pi_kpi}The $\overline{B}$ candidate mass spectra for the
final state
$D^{*+}\pi^+\pi^-\pi^-\pi^0$, with $D^0\to K^-\pi^+$ (a) for 
$\Delta E$
sidebands and (b) for $\Delta E$ consistent with zero. The curve 
in (a) is a
fit to the background distribution described in the text, while in 
(b) the
shape from (a) is used with the normalization allowed to float and 
a signal
Gaussian of width 2.7 MeV is added.}
\end{figure}

\subsection{Branching Fraction and $(4\pi)^-$ Mass Spectrum}

We start with the $D^0\to K^-\pi^+$ decay mode.
We show the candidate $\overline{B}$ mass distribution, $M_B$, for 
$\Delta E$ in the side-bands from -5.0 
to -3.0$\sigma$ and 3.0 to 5.0$\sigma$
on Fig.~\ref{bm_4pi_kpi}(a). The $\Delta E$ resolution
is 18 MeV ($\sigma).$ The sidebands give a good representation of the 
background in the signal region.
We fit this distribution with a shape given as 
\begin{equation}
\label{eq:background}
back(r)=p_1 r\sqrt{1-r^2}e^{-p_2(1-r^2)}~~~,
\end{equation}
where $r=M_B/5.2895$~GeV, and the $p_i$ are parameters given by the 
fit.

\begin{figure}[hbt]
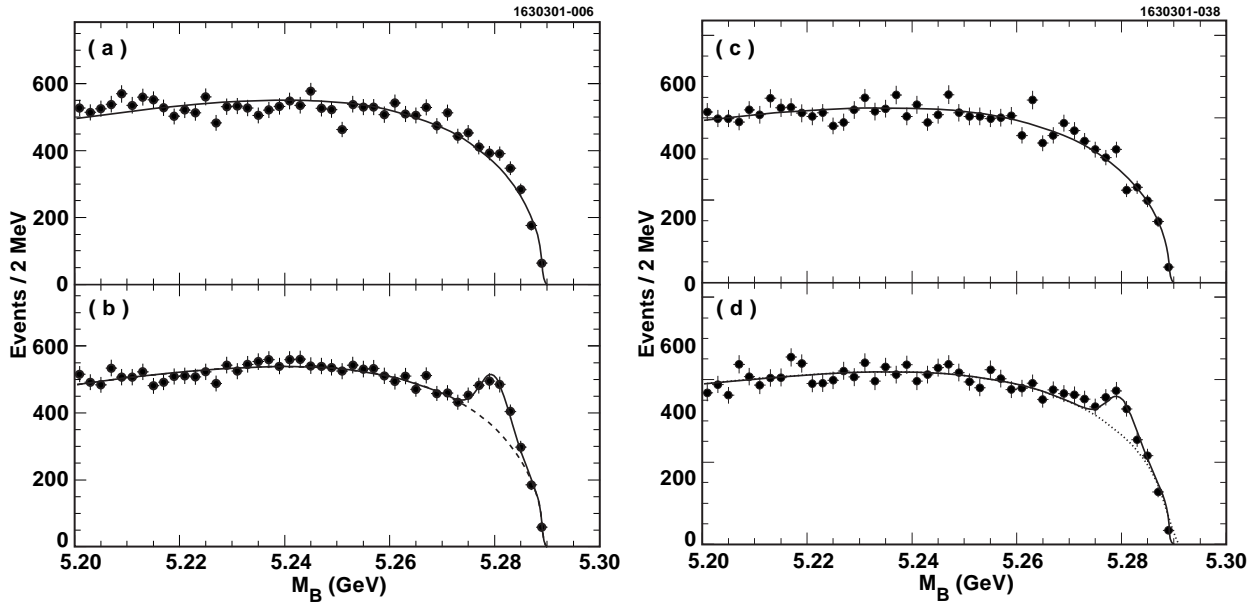

\centerline{\epsfig{figure=1630301-006.eps,height=3.2in}\hspace{-0.009in}
\epsfig{figure=1630301-038.eps,height=3.2in}}
\caption{\label{bm_4pi_k2pi}The $\overline{B}$ candidate mass spectra for
the final state $D^{*+}\pi^+\pi^-\pi^-\pi^0$, the left-side plots are
for $D^0\to K^-\pi^+\pi^0$ (a)  
$\Delta E$ sidebands, (b) for $\Delta E$ consistent with zero; the right-side
plots are for $D^0\to K^-\pi^+\pi^+\pi^-$ (c)  
$\Delta E$  sidebands, (d) for $\Delta E$ consistent with zero. The curves 
in the top plots, (a) and (c), are
fits to the background distribution described in the text, while in the 
bottom plots, (b) and (d), the
shapes from (a) and (c) are used with the normalization allowed to float and 
a signal Gaussian of width 2.7 MeV is added.}
\end{figure}

 We next view the $M_B$ distribution for events having  $\Delta E$ 
within 2$\sigma$ around zero in
Fig.~\ref{bm_4pi_kpi}(b). This distribution is fit with a Gaussian 
Signal function of width 2.7 MeV and the background function found above 
whose 
normalization is allowed to vary. The Gaussian signal width is found from
Monte Carlo simulation. The largest and dominant component results from the energy spread of the beam. We find 358$\pm$29 events in the 
signal peak.

We repeat this procedure for the other two $D^0$ decay modes. The 
$M_B$ spectrum for the $\Delta E$ sidebands and the signal region is shown 
in Fig.~\ref{bm_4pi_k2pi}. The $\Delta E$ resolution
is 22 MeV in the $K^-\pi^+\pi^0$ mode and 18 MeV in the 
$K^-\pi^+\pi^+\pi^-$ mode.
Signal to background ratios are worse in these two modes, but the
significance in both modes is quite large. The numbers of signal events are shown in Table~\ref{table:4pievents}.

\begin{table}[hbt]
\begin{center}
\caption{Event numbers for the $D^{*+}\pi^+\pi^-\pi^-\pi^0$ final 
state}
\label{table:4pievents}
\begin{tabular}{lc}
$D^0$ Decay Mode & Fitted \# of events \\\hline
$K^-\pi^+$&               358$\pm$29       \\
$K^-\pi^+\pi^0$ &          543$\pm$49            \\
$K^-\pi^+\pi^+\pi^-$ &       329$\pm$41               \\
\end{tabular}
\end{center}
\end{table}

\begin{figure}[hbt]
\centerline{\epsfig{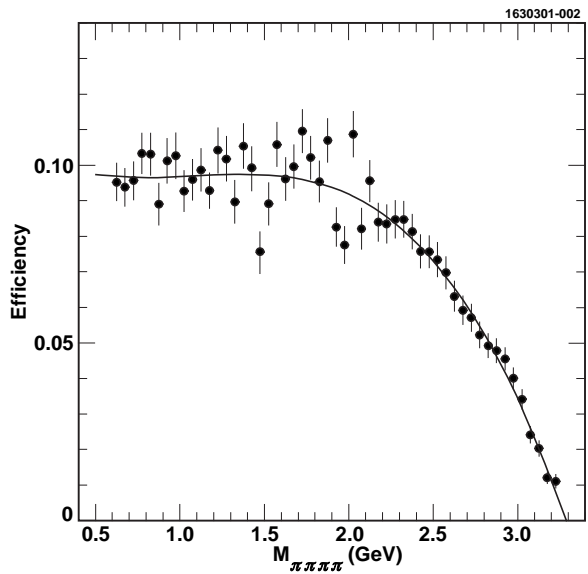}}
\vspace{0.3cm}
\caption{ \label{beff_ds4pi_kpi}The efficiency for the final state
$D^{*+}\pi^+\pi^-\pi^-\pi^0$, with $D^0\to K^-\pi^+$.}
\end{figure}

\begin{figure}[htb]
\centerline{\epsfig{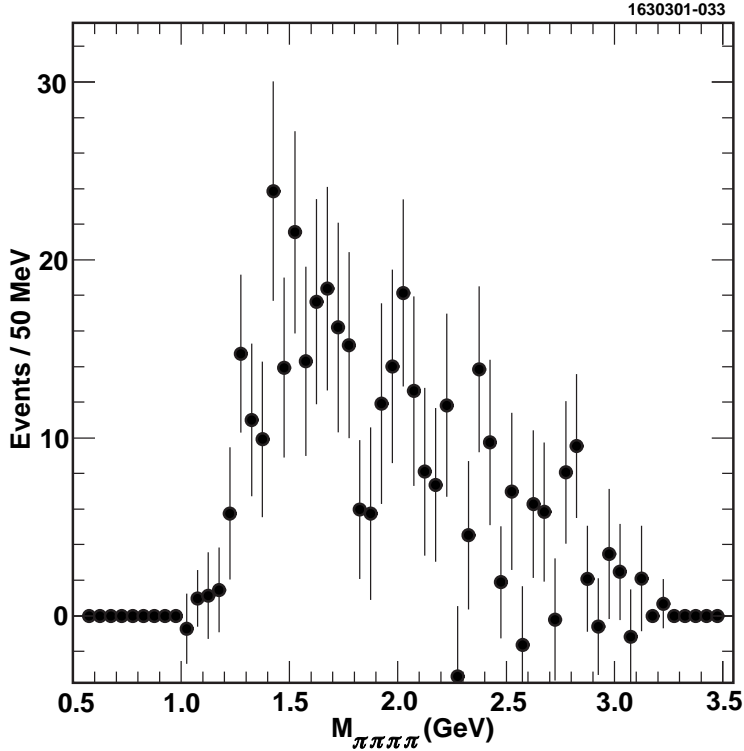}}
\caption{ \label{m4pi_ds4pi_kpi}The invariant mass spectra of 
$\pi^+\pi^-\pi^-\pi^0$ for the final state
$D^{*+}\pi^+\pi^-\pi^-\pi^0$, with $D^0\to K^-\pi^+$, found by 
fitting the $\overline{B}$ yield in bins of 4$\pi$ mass.}
\end{figure}

We choose to determine the branching fraction using only the 
$D^0\to K^-\pi^+$ 
decay mode because of the relatively large backgrounds in the 
other modes and
the decreased systematic error due to having fewer particles in 
the final state.
In order to find the branching ratio we use the Monte Carlo 
generated efficiency, shown in Fig.~\ref{beff_ds4pi_kpi} as a 
function of $(4\pi)^-$ mass. The efficiency falls off at larger 
$(4\pi)^-$ masses because the detection
of the slow $\pi^+$ from the $D^{*+}$ decay becomes increasingly 
more difficult.
Since the efficiency varies with mass we need to determine the 
$(4\pi)^-$ mass spectrum.
To rid ourselves of the problem of the background shape, we fit 
the $\overline{B}$ 
candidate mass spectrum in 50 MeV bins of $(4\pi)^-$ mass. (The 
mass resolution is approximately 12 MeV.) 
The resulting $(4\pi)^-$ mass 
spectrum is shown in Fig.~\ref{m4pi_ds4pi_kpi}. There are 
indications of a low-mass structure around
1.4 GeV, that will be investigated further in this paper.

We find
\begin{equation}
{\cal B}(\overline{B}^0\to D^{*+}\pi^+\pi^+\pi^-\pi^0)
=(1.72\pm 0.14\pm 0.24)\%~~~.
\end{equation}

The systematic error arises mainly from our lack of knowledge 
about the 
tracking and $\pi^0$ efficiencies. We assign errors of $\pm$2.2\% 
on the 
efficiency of each charged track, $\pm$5\% for the slow pion from 
the $D^{*+}$, and 
$\pm$5.4\% for the $\pi^0$. The error due to the 
background shape is evaluated 
in three ways. First of all, we change the background shape by 
varying the
fitted parameters by 1$\sigma$. This results in a change of 
$\pm$3\%. Secondly,
we allow the shape, $p_2$, to vary (the normalization, $p_1$, was 
already
allowed to vary). This results in 3.8\% increase in the number of 
events.
Finally, we choose a different background function
\begin{equation}
\label{eq:back2}
back'(r)=p_1 r\sqrt{1-r^2}\left(1+p_2 r + p_3 r^2 +p_4 r^3 
\right)~~~,
\end{equation}
and repeat the fitting procedure. This results in a 3.7\% 
decrease in the
number of events. Taking a conservative estimate of the systematic 
error due to
the background shape we arrive at $\pm$3.8\%. We use the current 
particle data group
values for the relevant $D^{*+}$ and $D^0$ branching ratios of
(68.3$\pm$1.4)\% ($D^{*+}\to\pi^+ D^0$) and 
(3.85$\pm$0.09)\% ($D^0\to K^-\pi^+$), respectively \cite{PDG}. 
The relative
errors, 2.0\% for the $D^{*+}$ branching ratio and 2.3\% for the 
$D^0$ are
added in quadrature to the background shape error, the $\pi^0$ detection efficiency uncertainty and the tracking error. The
total tracking error is found by adding the error in the charged particle
track finding efficiency linearly for the 5 ``fast" charged tracks and then in quadrature with the slow pion from the $D^{*+}$ decay. The
total systematic error is 14\%.

\begin{figure}[htbp]
\centerline{\epsfig{figure=1630301-031.eps,height=3.1in}
\epsfig{figure=1630301-028.eps,height=3.1in}}
\centerline{\epsfig{figure=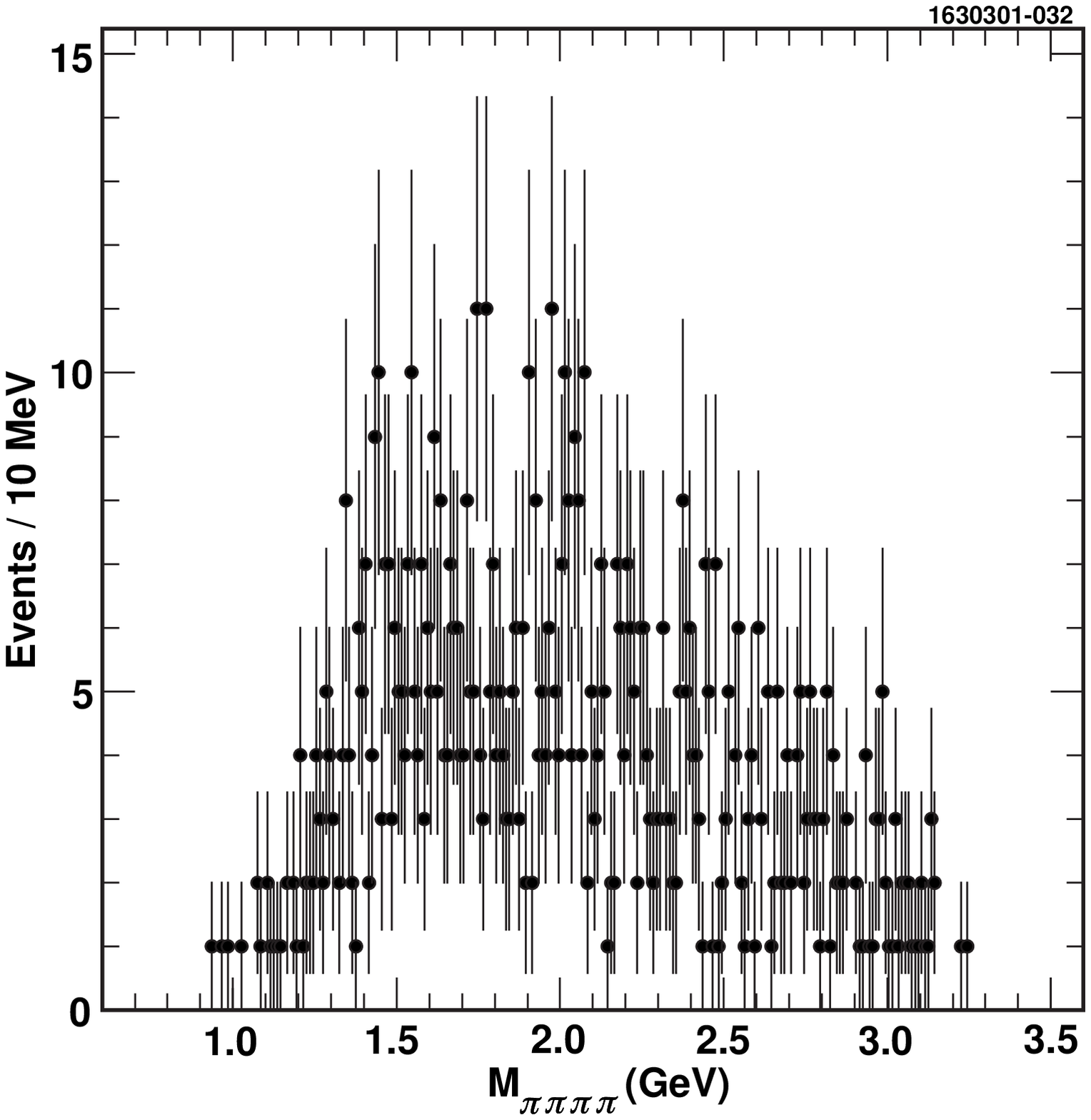,height=3.1in}
\epsfig{figure=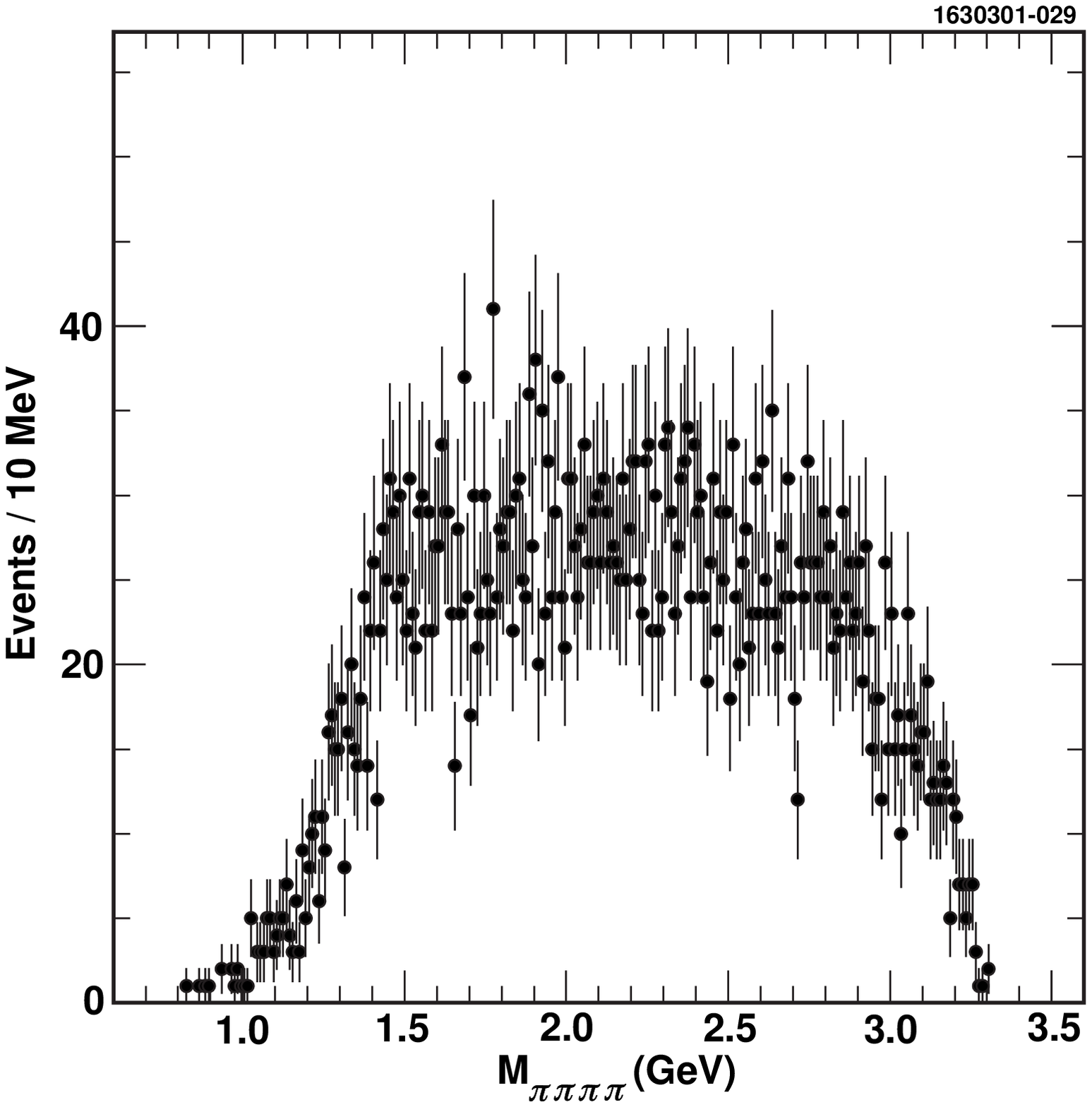,height=3.1in}}
\caption{\label{m4pi_c_all}The invariant mass spectra of 
$\pi^+\pi^-\pi^-\pi^0$ for the final state
$D^{*+}\pi^+\pi^-\pi^-\pi^0$, with $D^0\to K^-\pi^+$ (upper left), 
and the sum of all three $D^0$ decay modes (upper right). Events are selected by being
within $2\sigma$ of the $\overline{B}$ mass. The solid histogram is the background
estimate from the $M_B$ lower sideband and the dashed histogram is from the
$\Delta E$ sidebands; both are normalized to the fitted number of background
events. The same distributions in smaller bins (lower plots).}
\end{figure}

We wish to search for narrow structures. However, we cannot fit the $\overline{B}$ mass
spectrum in small $(4\pi)^-$ mass intervals due to a lack of statistics.
Thus we plot the  $(4\pi)^-$ mass for events
in the $M_B$ peak for the $D^0\to K^-\pi^+$ mode and the sum of all three modes in Fig.~\ref{m4pi_c_all}. We also plot two background samples: events at lower
$M_B$ (5.203 - 5.257 GeV) and those in the $\Delta E$ sideband separately.
First we view the plots in the canonical 50 MeV bins. Both background
distributions give a consistent if somewhat different estimates of the
background shape. (Each background distribution has been normalized
to the absolute number of background events as determined by the
fit to the $M_B$ distribution.) In any case no prominent narrow
structures appear in the histograms for the 10 MeV binning.

\begin{figure}[htb]
\centerline{\epsfig{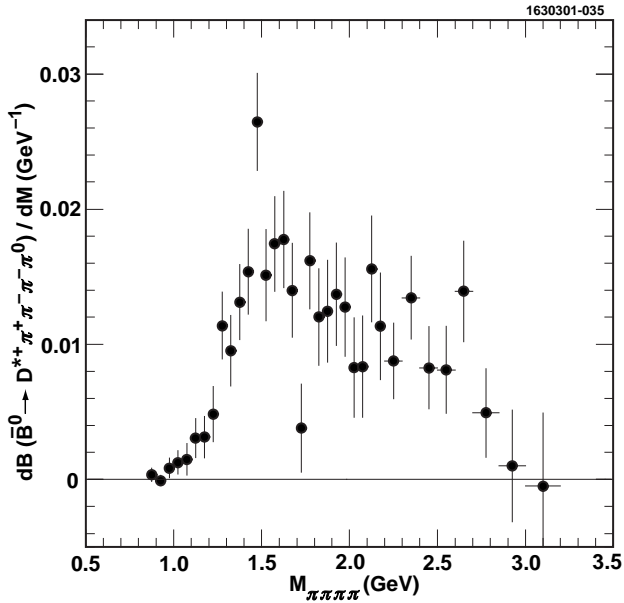}}
\caption{\label{m4pi_rebin}The efficiency-corrected background-removed
invariant mass spectra of $\pi^+\pi^-\pi^-\pi^0$ for the final state
$D^{*+}\pi^+\pi^-\pi^-\pi^0$, for  
the sum of all three $D^0$ decay modes. (There is an additional scale
uncertainty of 14\%.)}
\end{figure}

The most accurate distribution of $4\pi^-$ mass is obtained by using the
data in all three $D^0$ decay modes. The $4\pi^-$ mass distribution
shown in Fig.~\ref{m4pi_rebin} was found by fitting the $M_B$ candidate mass
distributions
summed together. The
distribution has been corrected for efficiency as a function of mass. There is
an additional 14\% systematic scale uncertainty on all the points.

\section{The $\overline{B}^0\to D^{*+}\omega\pi^-$ Decay}
\label{sec:Dspomegapi}
\begin{figure}[hbt]
\vspace{-.5cm}
\centerline{\epsfig{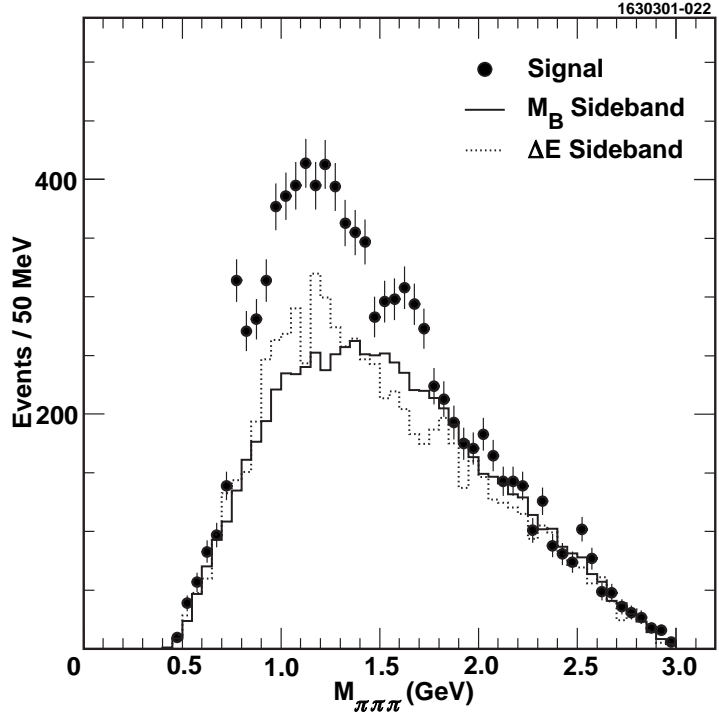}}
\caption{ \label{m3pi_c_all}The invariant mass spectra of 
$\pi^+\pi^-\pi^0$ for the final state
$D^{*+}\pi^+\pi^-\pi^-\pi^0$ for all three $D^0$ decay modes.
The solid histogram is the background
estimate from the $M_B$ lower sideband and the dashed histogram is 
from the
$\Delta E$ sidebands; both are normalized to the fitted number of 
background events. (There are two mass combinations per event.)}
\end{figure}

To investigate the composition of the $(4\pi)^-$ final state,
we now  investigate the $\pi^+\pi^-\pi^0$ mass spectrum for the
events in the $\overline{B}$ peak. All three $D^0$ decay modes are used. 
We show the $\pi^+\pi^-\pi^0$ invariant mass distribution for 
events
in the $\overline{B}$ mass peak in Fig.~\ref{m3pi_c_all} (there are two 
combinations per event). A clear signal is visible at the
$\omega$. The histograms on the figure are for events either in the lower 
$M_B$ range, from 5.203 GeV to 5.257 GeV, or in the previously defined
$\Delta E$ sidebands; no $\omega$ signal is visible.

The purity of the $\omega$ sample can be further improved by 
restricting candidates to certain regions of the 
Dalitz plot of the decay products. We define a cut on the Dalitz 
plot as follows.
Let $T_0$, $T_+$ and $T_-$ be the kinetic energies of the pions, 
and $Q$ be the difference between the $\omega$ mass, $M_{\omega}$
(equal to 782 MeV), 
and the mass of the 3 pions. We define two orthogonal coordinates 
$X$ and $Y$, where
\begin{eqnarray}
        X &= &3T_0/Q - 1       \\
        Y &= &\sqrt{3}(T_+ - T_-)/Q~~~.
\end{eqnarray}
The  kinematic limit that defines the Dalitz plot boundary is 
defined as
\begin{equation}
        Y_{boundary}^2 = {1\over 3} 
(X_{boundary}+1)(X_{boundary}+1+a) (1+b/(X_{boundary}+1-c))
\end{equation}
        where $a = 6m_0 / Q$,   $b = 6m^2 / (M_{\omega}Q)$, 
              $c = 3(M_{\omega}-m_0)^2/(2M_{\omega}Q)$, $m$ is the 
mass of a charged pion and $m_0$ the mass of the neutral pion.

For any set of three pion kinetic energies, we define a variable 
$r$, properly scaled to the kinematic limit as
\begin{equation}
r=\sqrt{{X^2+Y^2}\over{X_{boundary}^2+Y_{boundary}^2}}~~~,
\end{equation}
where the boundary values are found by following the radial vector 
from (0,0) through $(X,Y)$.

\begin{figure}[hbt]
\vspace{-0.3cm}
\centerline{\epsfig{figure=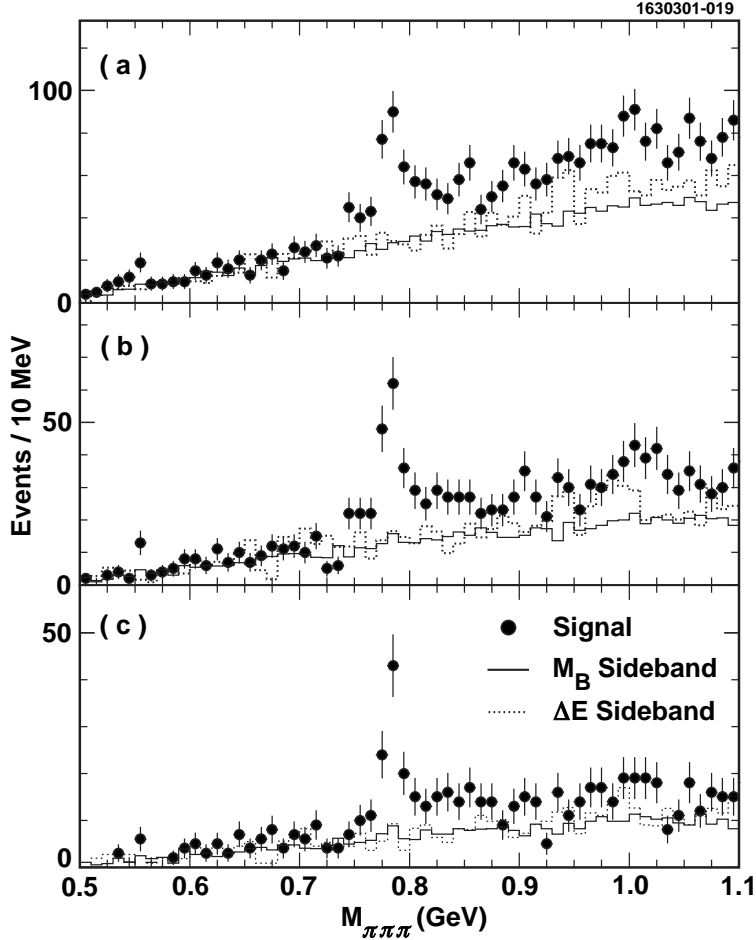,height=5.0in}}
\caption{ \label{m3_dal_all}The invariant mass spectra of 
$\pi^+\pi^-\pi^0$ for the final state
$D^{*+}\pi^+\pi^-\pi^-\pi^0$ for all three $D^0$ decay modes for
three selections on $r$ less than:  (a) 1, (b) 0.7 and (c) 0.5.
The solid histogram is the background
estimate from the $M_B$ lower sideband and the dashed histogram is 
from the
$\Delta E$ sidebands; both are normalized to the fitted number of 
background events.}
\end{figure}

For events in the $B$ mass peak we show in Fig.~\ref{m3_dal_all} 
the $\pi^+\pi^-\pi^0$ invariant mass for three different cuts on $r$. 
The $\omega$ signal is purified by restricting $r$, since the Dalitz plot density for a $1^-$ system decaying into $\pi^+\pi^-\pi^0$ peaks at $r$ equals zero \cite{Perkins}. 

\begin{figure}[htb]
\centerline{\epsfig{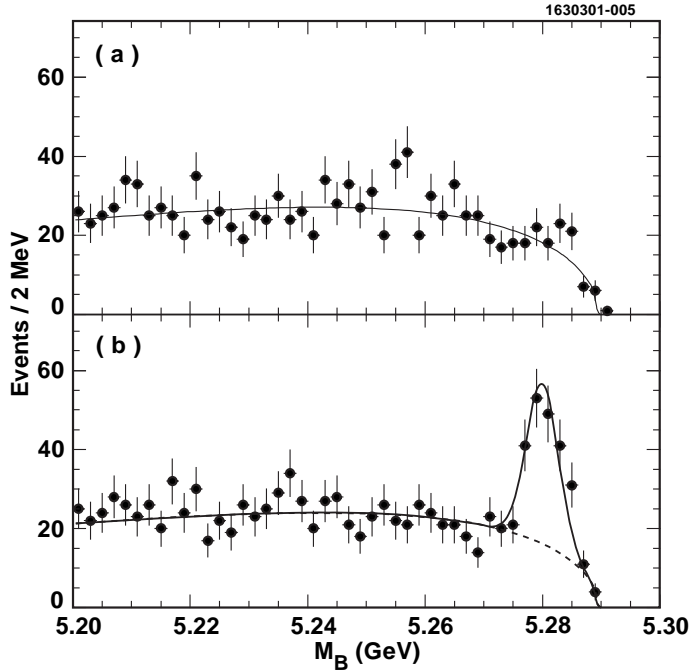}}
\vspace{0.4cm}
\caption{ \label{bm_4pi_all_4}The $M_B$ spectra for  
$D^{*+}\omega\pi^-$ for all three $D^0$ decay modes.
(a) $\Delta E$ sidebands and (b) $\Delta E$ around zero.}
\end{figure}

For further analysis we select $\omega$ candidates within the
$\pi^+\pi^-\pi^0$ mass window of 782$\pm$20 MeV with $r<0.7$. We 
abandon the $\chi_b^2$ cut here, as background is less of a problem. In 
Fig.~\ref{bm_4pi_all_4} we show the
$\overline{B}$ candidate mass distribution for the $D^{*+}\omega\pi^-$ final 
state summing
over all three $D^0$ decay modes. (The signal is fit with the same
prescription as before.) There are 136$\pm$15 events in the peak.
\begin{figure}[htb]
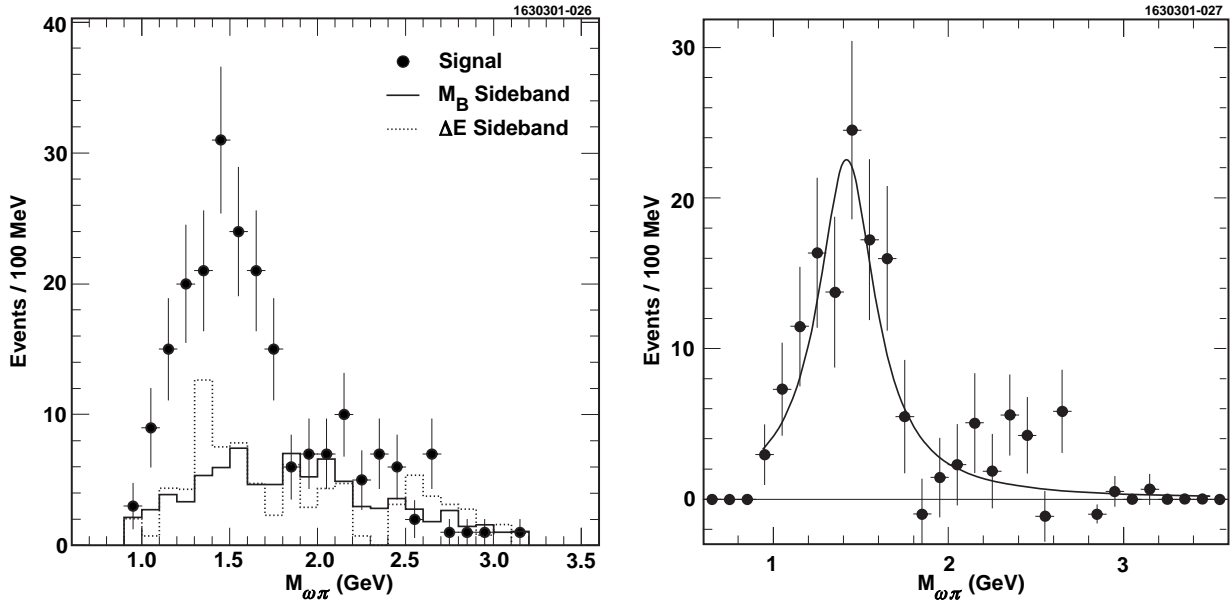

\centerline{\epsfig{figure=1630301-026.eps,height=3.2in}
\epsfig{figure=1630301-027.eps,height=3.2in}}
\caption{ \label{m4pi_c4_all}The invariant mass spectra of 
$\omega\pi^-$ for the final state
$D^{*+}\pi^+\pi^-\pi^-\pi^0$ for all three $D^0$ decay modes.
(left) Signal events
satisfy cuts on $B$ mass, $\Delta E$ and
$\omega$ mass (see text). The solid histogram is the background
estimate from the $M_B$ lower sideband and the dashed histogram is 
from the
$\Delta E$ sidebands; both are normalized to the fitted number of 
background
events. (right) The mass spectrum determined from fitting the 
$M_B$ distribution
and fit to a non-relativistic Breit-Wigner function that gives a mass
of 1432$\pm$37 MeV and a width of 376$\pm$47 MeV.}
\end{figure}	

In Fig.~\ref{m4pi_c4_all} we show
the $\omega\pi^-$ mass spectrum in the left-side plot. The solid 
histogram shows events from the 
lower $M_B$ sideband region suitably normalized. The dotted 
histogram shows 
the background estimate from the $\Delta E$ sidebands, again 
normalized. 
In the signal distribution there is a wide structure around 
1.4 GeV, that is inconsistent with background. We re-determine the 
$\omega\pi^-$
mass distribution by fitting the $M_B$ distribution in bins of 
$\omega\pi^-$ mass, and this is shown on the right-side.  

Knowing the $\omega\pi^-$ mass dependence of the efficiency we evaluate the 
branching fraction:
\begin{equation}
{\cal B}(\overline{B}^0\to D^{*+}\omega\pi^-)=(0.29 \pm 0.03 \pm 
0.04)\%~~~.
\end{equation}

We provisionally label the state at 1432 MeV the $A^-$ and 
investigate its
properties later. The $\omega\pi^-$ comprises about 17\% of the 
$(4\pi)^-$ final state.
All of the $\omega\pi^-$ final state is consistent with coming 
from $A^-$ decay.

\section{Observation of ${B}^-\to D^{*0}\pi^+\pi^-\pi^-\pi^0$}
\label{sec:Ds04pi}
\begin{figure}[bht]
\vspace{-1.5cm}
\centerline{\epsfig{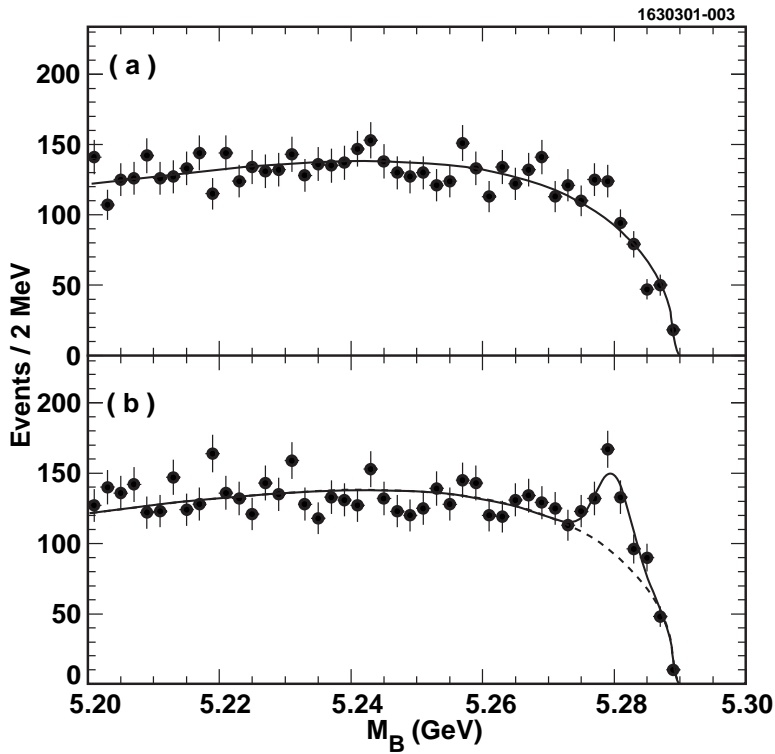}}
\caption{ \label{bm_4pi_0kpi}The $\overline{B}$ candidate mass spectra for 
the final state
$D^{*0}\pi^+\pi^-\pi^-\pi^0$, with $D^0\to K^-\pi^+$. Signal events
satisfy cuts on $B$ mass, $\Delta E$ and
$\omega$ mass (see text).  (a) for 
$\Delta E$
sidebands and (b) for $\Delta E$ consistent with zero. The curve 
in (a) is a
fit to the background distribution described in the text, while in 
(b) the
shape from (a) is used with the normalization allowed to float and 
a signal
Gaussian of width 2.7 MeV is added.}
\end{figure}

We proceed in the same manner as for the $\overline{B}^0$ reaction 
with the exception that we use the $D^{*0}\to \pi^0 D^0$ decay 
mode and 
restrict ourselves to the $D^0\to K^-\pi^+$ decay mode only due to 
large backgrounds
in the other modes. The $\chi^2$ is calculated according to 
equation~\ref{eq:chisq} and we use a cut value of 8.
The $M_B$ distributions for $\Delta E$ sidebands and signal data are 
shown in 
Fig.~\ref{bm_4pi_0kpi} for the $D^0\to K^-\pi^+$ decay mode. The $\Delta E$
resolution is 18 MeV. We 
see a signal of 195$\pm$26 events yielding a branching fraction of 
\begin{equation}
{\cal{B}}({B}^-\to D^{*0}\pi^+\pi^-\pi^-\pi^0) = (1.80\pm 0.24\pm 
0.27) \%~~~~.
\end{equation}

\begin{figure}[thb]
\centerline{\epsfig{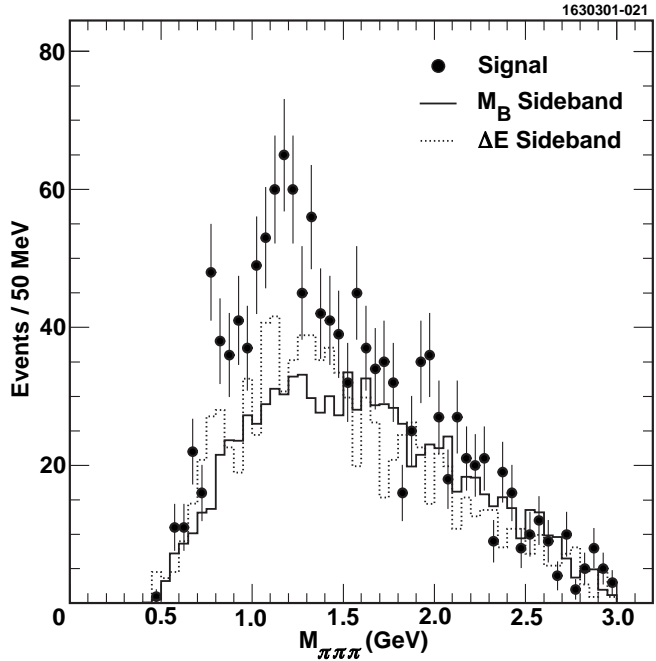}}
\vspace{0.2cm}
\caption{ \label{m3pi_c_0kpi}The invariant mass spectra of 
$\pi^+\pi^-\pi^0$ for the final state
$D^{*0}\pi^+\pi^-\pi^-\pi^0$ for the $D^0\to K^-\pi^+$ decay 
mode.
The solid histogram is the background
estimate from the $M_B$ lower sideband and the dashed histogram is 
from the
$\Delta E$ sidebands; both are normalized to the fitted number of 
background events. There are two combinations per event.}
\end{figure}

The $\pi^+\pi^-\pi^0$ mass spectrum shown in 
Fig.~\ref{m3pi_c_0kpi} shows the presence of an $\omega$.
Selecting on the presence of an $\omega$ with $r<0.7$ we show the 
sideband and signal plots in Fig.~\ref{bm_4pi_0kpi_4}.
(Here we do not use the previously defined $\chi^2$ cut.)
\begin{figure}[htb]
\centerline{\epsfig{figure=1630301-004.eps,height=3.6in}}
\caption{ \label{bm_4pi_0kpi_4}The $M_B$ spectra for  
$D^{*0}\omega\pi^-$ for the $D^0\to K^-\pi^+$ decay mode.
(a) $\Delta E$ sidebands and (b) $\Delta E$ around zero.}
\end{figure}
The branching ratio, based on 26$\pm$6 events is
\begin{equation}
{\cal{B}}({B}^-\to D^{*0}\omega\pi^-) = (0.45\pm 0.10 \pm 0.07) 
\%~~~~.
\end{equation}

In Fig.~\ref{m4pi_c4_0kpi} we show the $\omega\pi^-$
mass spectrum.
We see an enhancement at around 1.4 GeV as in the
neutral $\overline{B}$ case. (We do not have enough statistics here 
to fit the $M_B$ distribution in bins of $\omega\pi^-$ mass.)
The $\omega\pi^-$ fraction of the 
$(4\pi)^-$ final 
state is 25\%, and all the $\omega\pi^-$ is consistent with coming 
from the $A^-$.

\begin{figure}[htb]
\centerline{\epsfig{figure=1630301-024.eps,height=3.2in}
\epsfig{figure=1630301-025.eps,height=3.2in}}
\caption{ \label{m4pi_c4_0kpi}The invariant mass spectra of 
$\omega\pi^-$ for the final state
$D^{*0}\pi^+\pi^-\pi^-\pi^0$ for the $D^0\to K^-\pi^+$ decay mode.
(left) Signal events
satisfy cuts on $B$ mass, $\Delta E$ and
$\omega$ mass (see text). The solid histogram is the background
estimate from the $M_B$ lower sideband and the dashed histogram is 
from the
$\Delta E$ sidebands; both are normalized to the fitted number of 
background
events. (right) The data fit to a non-relativistic Breit-Wigner signal and a
smooth background function.
The mass and width are 1367$\pm$75 MeV and 439$\pm$135 MeV, respectively.}
\end{figure}

\section{Analysis of $D^{*+}\omega\pi^-$ Decay Angular Distributions}
\label{sec:angular}

 The $A^-$ is produced along with a spin-1 $D^*$ from a spin-0 
$\overline{B}$. 
If the $A^-$ is spin-0 the $D^*$ would be fully polarized in the 
$(J,~J_z)= (1,0)$ state.
If the $A^-$ were to be spin-1 any combinations of z-components 
would be
allowed. It is natural then to examine the helicity angle of the 
$D^{*+}$ by
viewing the cosine of the helicity angle of the $\pi^+$ with 
respect to the $\overline{B}$ in the $D^{*+}$ rest frame.

Another decay angle that can be examined is that of the 
$\omega\pi$ system.
If the $A^-$ is spin-0, the $\omega$ is polarized in the (1,0) 
state
and may be if the $A^-$ is spin-1. Here the helicity angle is 
defined as the
angle between the normal to the $\omega$ decay plane and the 
direction of the
$A^-$ in the $\omega$ rest frame. For a spin-0 $A^-$ the 
distribution will be cosine-squared. Again full polarization is
possible if the $A^-$ is other 
than spin-0, but any distribution other than cosine-squared would 
demonstrate that the spin is not equal to zero.

For this analysis we use all three $D^0$ final states for the 
$D^{*+}$ final 
state. To find the distributions we fit the number of events
in the $M_B$ candidate plot selected on different angle bins. The 
$\omega\pi$ mass is required to be between 1.1 and 1.9 GeV. This restriction leaves 111$\pm$13 events.

\begin{figure}[bht]
\centerline{\epsfig{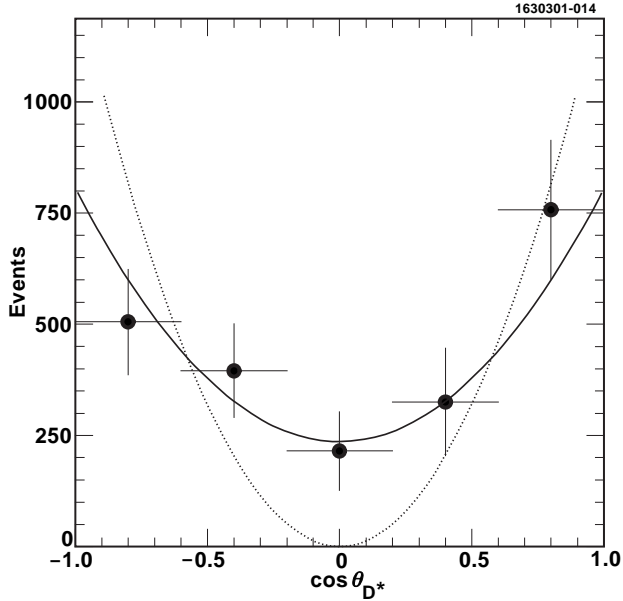}}
\caption{ \label{cosd} The cosine of the angle between the $D^0$ 
and the
$D^{*}$ flight direction in the $D^{*}$ rest frame for the 
$D^{*}A^-$ final
state (solid points) along with a fit (solid curve) allowing the
amount of longitudinal and traverse polarization to vary. The dotted curve
is the expectation for a spin-0 $A^-$.}
\end{figure}

In Fig.~\ref{cosd} we show the helicity angle distribution, 
$\cos\theta_{D^*}$ 
for the $D^{*}$ decay. The data have been corrected for acceptance.
We also show the expectation for 
spin-0 from the Monte Carlo. The data have been fit for the fraction
of longitudinal polarization. We find 
\begin{equation}
{\Gamma_L \over \Gamma}=0.63\pm 0.09 ~~.
\end{equation}
The systematic error is much smaller than the statistical error. 

\begin{figure}[htb]
\centerline{\epsfig{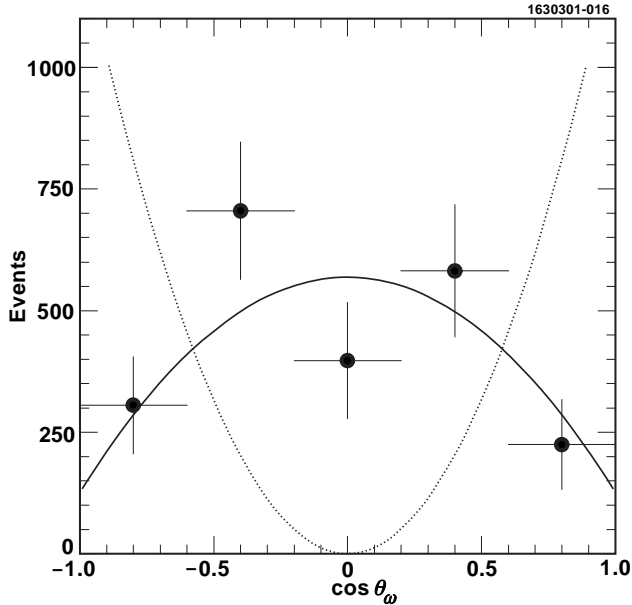}}
\caption{\label{cosom} The cosine of the angle between the normal 
to the
$\pi^+\pi^-\pi^0$ decay plane and the $\omega$ boost direction
 for the $D^{*}A^-$ final state (solid points) along with a fit (solid curve)
 allowing the
amount of longitudinal and traverse polarization to vary. The dotted
curve is the expectation for a spin-0 $A^-$.}
\end{figure}
The $\cos\theta_{D^*}$ distribution is not consistent with full 
polarization, yielding a $\chi^2$ of 17.7 for 5 degrees of freedom. 
The helicity angle
distribution for the $A^-\to\omega\pi^-$, $\cos\theta_{\omega}$, 
is shown on Fig.~\ref{cosom}. Furthermore 
the $\cos\theta_{\omega}$ distribution is quite inconsistent with 
a $\cos^2\theta_{\omega}$, yielding a $\chi^2$ of 109 for 5 degrees 
of freedom. Therefore, we rule out a spin-0 assignment for the $A^-$.

To determine the $J^P$ we need a more well defined final state. 
This is provided by analysis of $\overline{B}\to D\omega\pi^-$ decays.

\section{Observation of $\overline{B}\to D\omega\pi^-$ Decays}
\label{sec:Domegapi}  
\subsection {$\overline{B}$ candidate selection}  
  
Here we study the reactions $\overline{B}\to D\omega\pi^-$, with 
either a 
$D^0\to K^-\pi^+$ or $D^+\to K^-\pi^+\pi^+$ decay. Other $D^0$ or 
$D^+$ decays have substantially larger backgrounds. 
    
Although we are restricting our search to $\omega$'s, we define 
two $\pi^+\pi^-\pi^0$
samples. One within 20 MeV of the known $\omega$ mass (782 MeV) 
and the other in 
either low mass or high mass sideband defined as three $\pi$ mass 
either between
732 and 752 MeV or between 812 and 832 MeV. We also require a cut 
on the $\omega$
Dalitz plot of $r < 0.7.$

To reduce backgrounds we define 
\begin{equation}\label{eq:chisqD} 
\chi_b^2=  
\left({{\Delta M_{D}}\over {\sigma(\Delta M_{D})}}\right)^2 +  
\left({{\Delta M_{\omega}}\over {\sigma(\Delta 
M_{\omega})}}\right)^2 +  
\left({{\Delta M_{\pi^0}}\over {\sigma(\Delta 
M_{\pi^0})}}\right)^2~~~, 
\end{equation} 
where $\Delta M_{D}$ is the invariant candidate $D^0$ mass minus 
the known $D^0$
mass, $\Delta M_{\omega}$ is the invariant candidate $\omega$ mass 
minus the known
$\omega$ mass, and $\Delta M_{\pi^0}$ is the measured 
$\gamma\gamma$ invariant
mass minus the known $\pi^0$ mass. The $\sigma$'s are the 
measurement errors.
We select candidate events requiring that  $\chi^2_b$ is
$<$ 12 for the $K\pi$ mode and $<$6 for the $K\pi\pi$ mode. 

\subsection{$B^-\to D^0\omega\pi^-$ Signal}

\begin{figure}[htb] 
\centerline{\epsfig{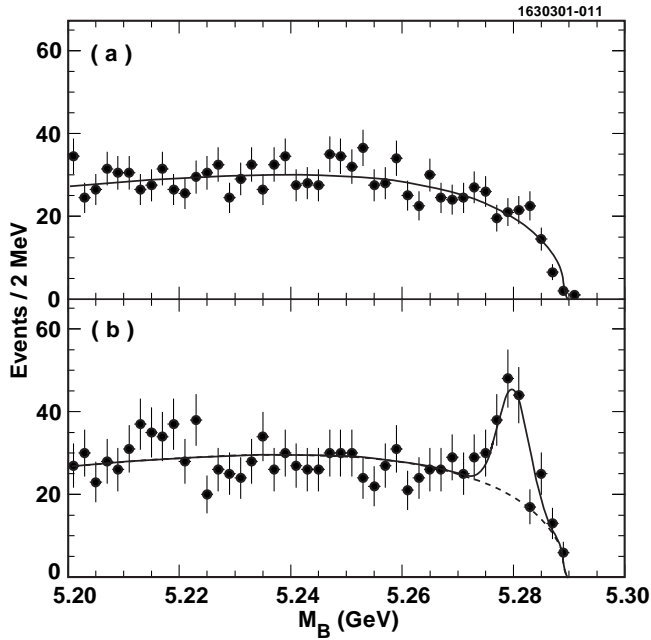}} 
\caption{ \label{bm_omega_D0}The $\overline{B}$ candidate mass spectra for 
the final state 
$D^{0}\omega\pi^-$, with $D^0\to K^-\pi^+$. (a) for $\Delta E$ 
sidebands, and (b) for $\Delta E$ consistent with zero. The 
vertical scale
in (a) was multiplied by 0.5 to facilitate comparison.  The curve 
in (a) is a 
fit to the background distribution described in the text, while in 
(b) the 
shape from (a) is used with the normalization allowed to float and 
a signal Gaussian of width 2.7 MeV is added.} 
\end{figure}

We start with the $D^0\to K^-\pi^+$ decay mode, for events in the 
$\omega$ peak. 
We show the candidate $\overline{B}$ mass distribution, $M_B$, for
$\Delta E$ in the side-bands from -7.0 
to -3.0$\sigma$ and 7.0 to 3.0$\sigma$ 
on Fig.~\ref{bm_omega_D0}(a).  This gives a good representation of the 
background in the signal region. The $\Delta E$ resolution 
is 18 MeV ($\sigma).$ 
We fit this distribution with the shape given in
Equation~\ref{eq:background}.

We next view the $M_B$ distribution for events having  $\Delta E$  
within 2$\sigma$ around zero in 
Fig.~\ref{bm_omega_D0}(b). This distribution is fit with a 
Gaussian signal  
function of width 2.7 MeV and the background function found above 
whose normalization is allowed to vary. We find 88
$\pm$14 events in the signal peak.  
 
\begin{figure}[htb] 
\centerline{\epsfig{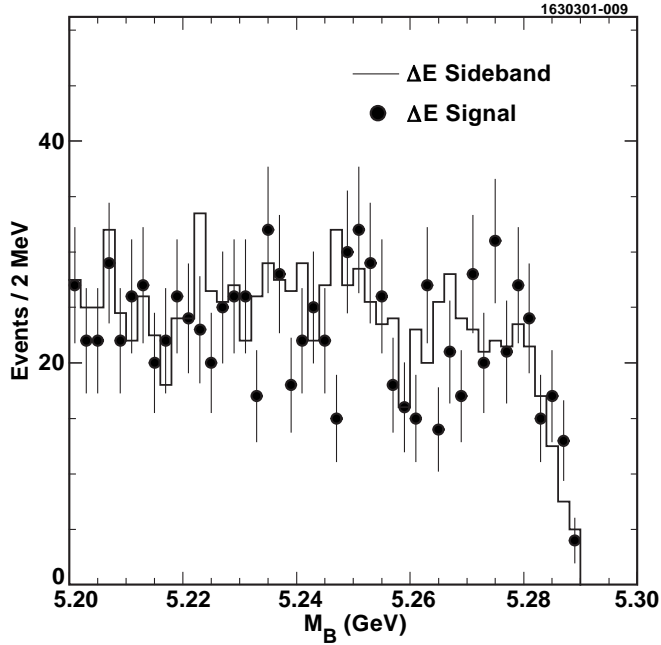}} 
\caption{ \label{bm_not_omega_D0}The $\overline{B}$ candidate mass spectra 
for the final state 
$D^{0}\omega\pi^-$, with $D^0\to K^-\pi^+$ and $\omega$ sidebands
for $\Delta E$ sidebands (histogram) and $\Delta E$ consistent 
with zero (points). The $\Delta E$ sideband numbers have been
divided by 2.} 
\end{figure}

We repeat this procedure for events in the $\omega$ sidebands. We 
use for
our $\chi^2_b$ definition pseudo-$\omega$ masses in the center of the sideband
intervals.
We show the $M_B$ distribution for events in the $\Delta E$ 
sideband,
defined above, and those having  $\Delta E$  within 2$\sigma$ 
around zero in 
Fig.~\ref{bm_not_omega_D0}. We find no significant signal.

\subsection{$\overline{B}^0\to D^+\omega\pi^-$ Signal} 
 
\begin{figure}[htb] 
\centerline{\epsfig{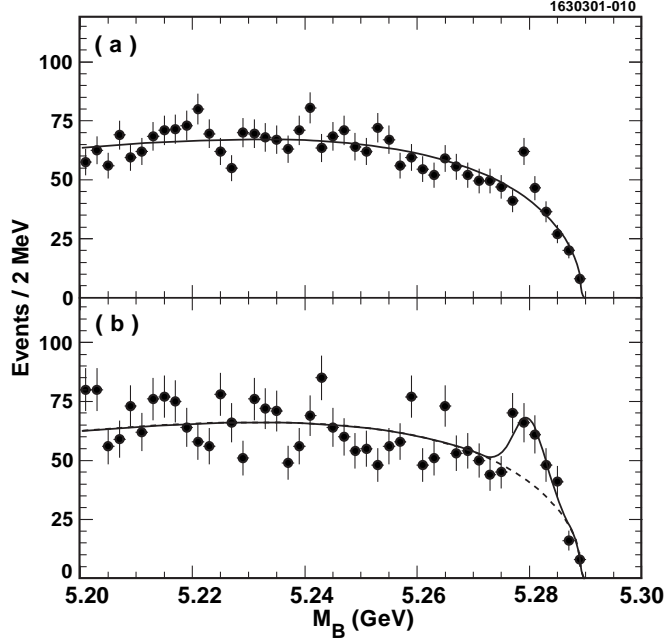}}
\vspace{0.1cm} 
\caption{ \label{bm_omega_D+}The $\overline{B}$ candidate mass spectra for 
the final state 
$D^{+}\omega\pi^-$, with $D^+\to K^-\pi^+\pi^+$ (a) for $\Delta E$ 
sidebands and (b) for $\Delta E$ consistent with zero. The 
vertical scale
in (a) was multiplied by 0.5 to facilitate comparison. The curve 
in (a) is a 
fit to the background distribution described in the text, while in 
(b) the 
shape from (a) is used with the normalization allowed to float and 
a signal Gaussian of width 2.7 MeV is added.} 
\end{figure}

The same procedure followed for the $D^0$ final state is used for the $D^+$
final state.  
We show the candidate $\overline{B}$ mass distribution, $M_B$, for events in the  
$\Delta E$ side-band
 on Fig.~\ref{bm_omega_D+}(a). The $\Delta E$ resolution 
is 18 MeV ($\sigma).$ This gives a good representation of the 
background in the signal region.  
We fit this distribution with a shape given in equation~\ref{eq:background}. 

We next view the $M_B$ distribution for events having  $\Delta E$  
within 2$\sigma$ around zero in Fig.~\ref{bm_omega_D+}(b). 
This distribution is fit with a Gaussian signal  
function of width 2.7 MeV and the background function found above 
whose  
normalization is allowed to vary. We find 91$\pm$18 events in the 
signal peak.  
 
\begin{figure}[htb] 
\centerline{\epsfig{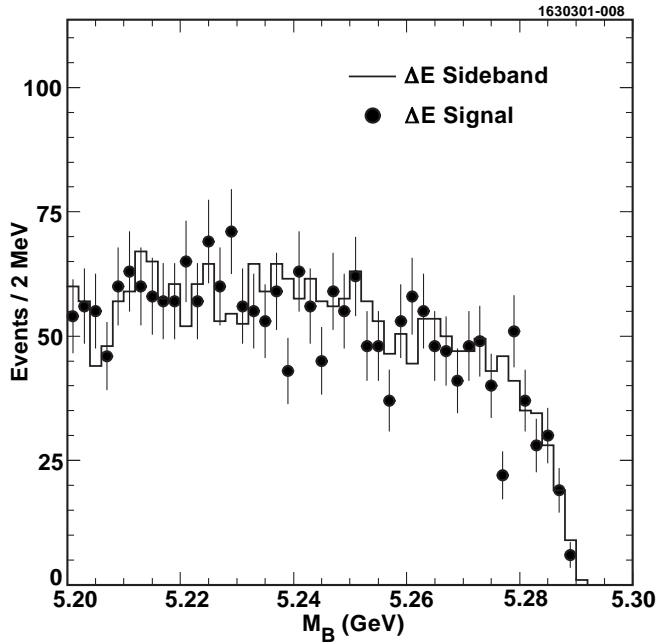}}  
\caption{ \label{bm_not_omega_D+}The $\overline{B}$ candidate mass spectra 
for the final state 
$D^{+}\omega\pi^-$, with $D^+\to K^-\pi^+\pi^+$ and $\omega$ 
sidebands
for $\Delta E$ sidebands (histogram) and $\Delta E$ consistent 
with zero (points). The $\Delta E$ sideband numbers have been
divided by 2.}
\end{figure}

We repeat this procedure for events in the $\omega$ sidebands. 
We show the $M_B$ distribution for both $\Delta E$ sidebands and  
$\Delta E$  
within 2$\sigma$ around zero in Fig.~\ref{bm_not_omega_D+}.

There is no evidence of any signal in the $\omega$ sideband plot, 
leading to the conclusion that the signal is associated purely 
with $\omega$.

\subsection{Branching Fractions}

We determine the branching ratios, shown in 
Table~\ref{table:Domgpievents},
by performing a Monte Carlo simulation of the
efficiencies in the two modes.
We use the current particle data group 
values for the relevant $\omega$, $D^{+}$ and $D^0$ branching 
ratios of 
(88.8$\pm$0.7)\% ($\omega\to \pi^+\pi^-\pi^0$), 
(9.0$\pm$0.6)\% ($D^{+}\to K^-\pi^+\pi^+$) and  
(3.85$\pm$0.09)\% ($D^0\to K^-\pi^+$) \cite{PDG}.
 The efficiencies listed in the table do not include these 
branching ratios \cite{spineff}.

\begin{table}[hbt]  
\begin{center}  
\caption{Branching Fractions for the $D\omega\pi^-$ final state}  
\label{table:Domgpievents}  
\begin{tabular}{lccc} 
$D$ Decay Mode & Fitted \# of events& Efficiency &Branching 
Fraction (\%)\\\hline  
$K^-\pi^+$&               88$\pm$14   & 0.064 &  
0.41$\pm$0.07$\pm$0.06   \\  
$K^-\pi^+\pi^+$ &         91$\pm$18   & 0.046 &  
0.28$\pm$0.05$\pm$0.04   \\
\end{tabular}  
\end{center}  
\end{table}

The systematic error arises mainly from our lack of knowledge 
about the  
tracking and $\pi^0$ efficiencies. We assign errors of $\pm$2.2\% 
on the  
efficiency of each charged track,  and  
$\pm$5.4\% for the $\pi^0$. The error due to the 
background shape is evaluated  
in three ways. First of all, we change the background shape by 
varying the 
fitted parameters by 1$\sigma$. This results in a change of 
$\pm$5.0\%. Secondly, 
we allow the shape, $p_2$, to vary (the normalization, $p_1$, was 
already 
allowed to vary). This results in 5.5\% increase in the number of events. 
Finally, we choose a different background function given in equation~\ref{eq:back2} 
and repeat the fitting procedure. This results in a 1.0\% decrease 
in the 
number of events. Taking a conservative estimate of the systematic 
error due to 
the background shape we arrive at $\pm$5.5\%.

\subsection {The $\omega\pi^-$ System} 
 
For all subsequent discussions we add the $D^0$ and $D^+$ final 
states together. 
We select sample of $\omega$'s in the $\pi^+\pi^-\pi^0$ mass 
window of 782$\pm$20 MeV using only combinations having $r<0.7$ 
in the Dalitz plot. 
 
\begin{figure}[htb]
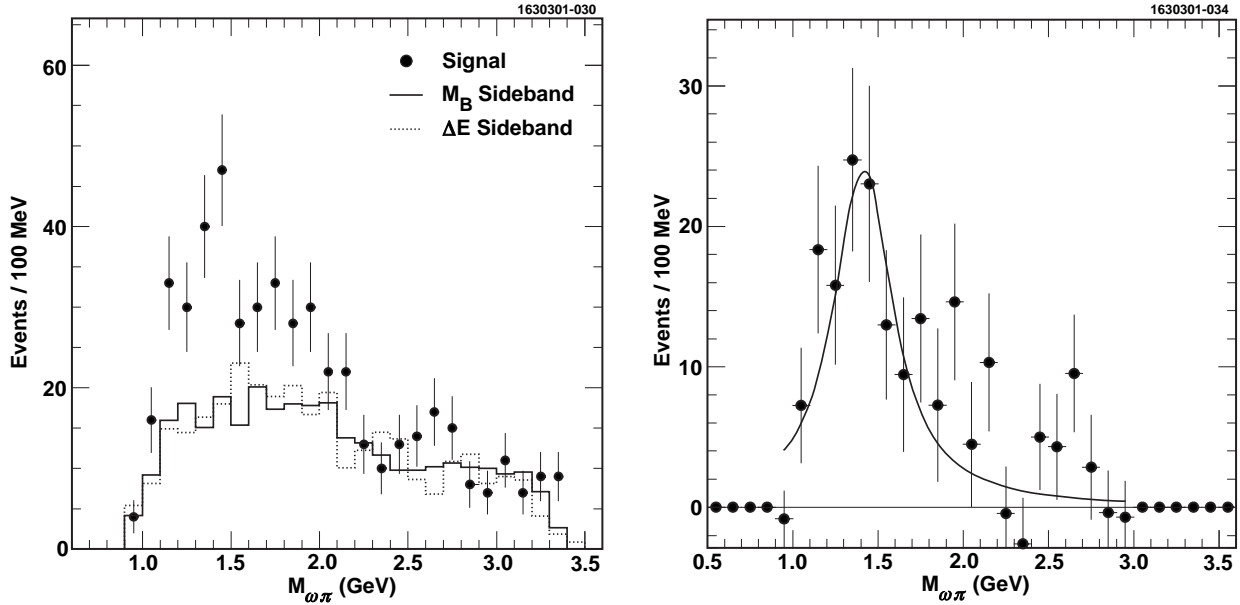
 
\centerline{ \epsfig{figure=1630301-030.eps,height=3.2in}
             \epsfig{figure=1630301-034.eps,height=3.2in} }
\caption{ \label{m_omega_pi} The invariant mass spectra of 
${\omega\pi^-}$ for the
final state $D\omega\pi^-$ for both ${D}$ decay modes. (left) The 
solid histogram
is the background estimate from the $M_B$ lower sideband and the 
dashed histogram
is from the $\Delta E$ sidebands; both are normalized to the 
fitted number of
background events. (right) The mass spectrum determined from 
fitting the $M_B$ 
distribution and fit to a Breit-Wigner function.
We find a peak value of 1415$\pm$43 MeV and a width of 419$\pm$110 MeV.}
\end{figure}

In Fig.~\ref{m_omega_pi} we show the $\omega\pi^-$ mass spectrum 
in the left-side plot. 
The solid histogram shows events from the  
lower $M_B$ sideband region (5.203 - 5.257 GeV) suitably 
normalized. The dotted histogram shows  
the background estimate from the $\Delta E$ sidebands, again 
normalized.  
In the signal distribution there is a wide structure around   
1.4 GeV, that is inconsistent with background. We re-determine the 
$\omega\pi^-$ 
mass distribution by fitting the $M_B$ distribution in bins of 
$\omega\pi^-$ mass, and this is shown on the right-side.  

This structure appears identical to the one we observed in
$\overline{B}\to D^{*}\omega\pi^-$ decays.
   
\subsection{Angular Distributions in $D\omega\pi^-$}
\label{sec:Dangular}
We can determine the spin and parity of the $A^-$ 
particle by
studying the angular distributions characterizing its decay 
products. 
The decay chain that we are considering is $B\rightarrow A\ D$; 
$A\rightarrow
\omega \pi$ and $\omega \rightarrow \pi ^+ \pi ^- \pi ^0$. The 
helicity formalism \cite{helicity} is 
generally used in the analysis of these 
sequential decays. This formalism is well suited to relativistic
problems involving particles with spin $\vec{J}$ and momentum 
$\vec{p}$ because the helicity operator $h=\vec{J}\cdot \vec{p}$ 
is invariant under both rotations and boosts along $\hat{p}$. 

\begin{figure}
\centerline{\epsfxsize=3.5in\epsffile{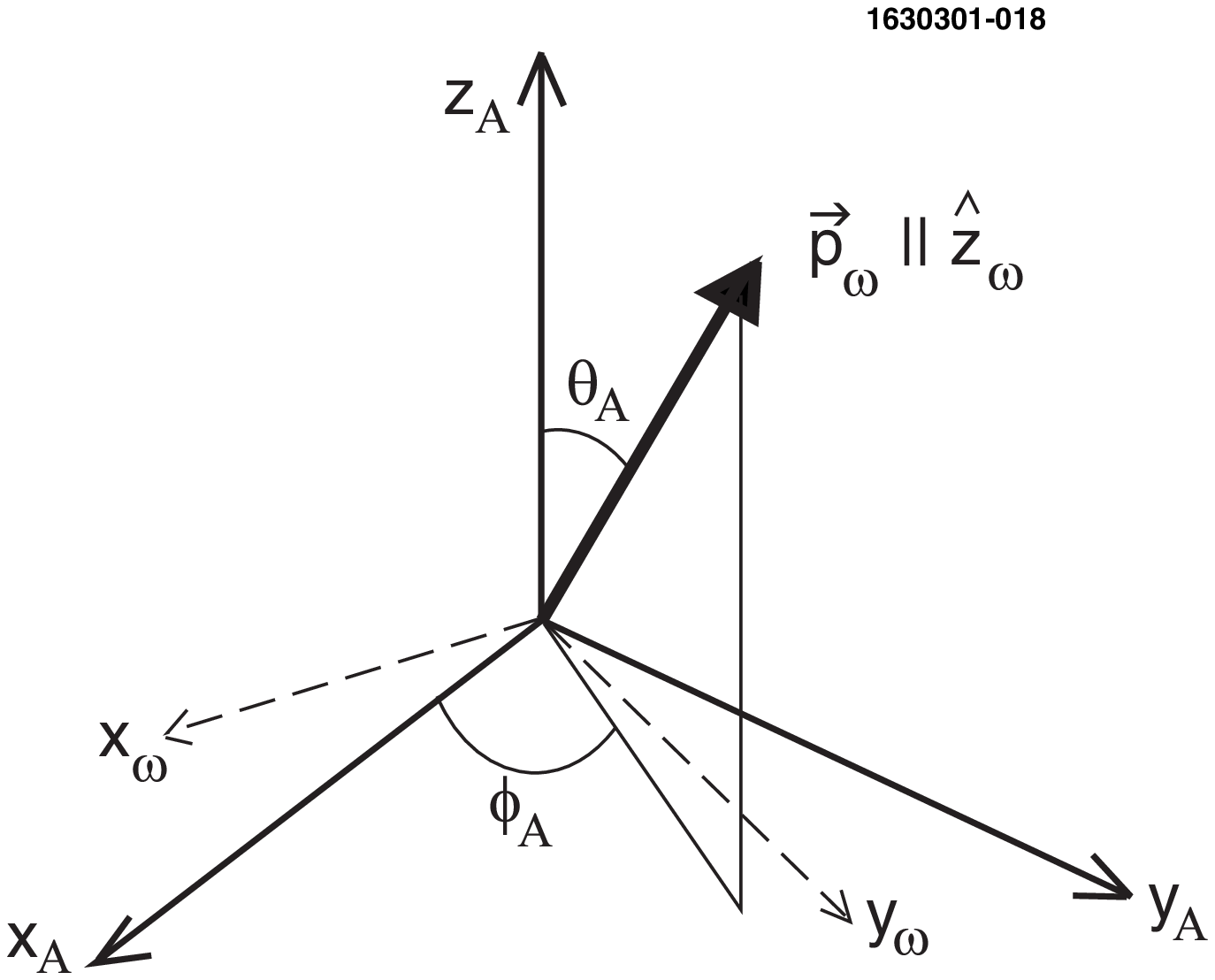}}
\vspace{2mm}
\caption{Relationship between the $A$ rest frame $x_Ay_Az_A$ and 
the $\omega$
rest frame $x_{\omega}y_{\omega}z_{\omega}$. $x_A$ and 
$x_{\omega}$ lie in 
the same plane.}
\label{eul-angles}
\end{figure} 
There are two relevant reference frames. The first one, that we 
will 
define $x_Ay_Az_A$ is the rest frame of the $A$ particle, with the
$\hat{z_A}$ axis 
pointing in the $A$ direction of motion in the $B$ rest frame. 
The $\hat{x_A}$ direction is arbitrary. The 
second
one, $x_{\omega}y_{\omega}z_{\omega}$, is
related to $x_Ay_Az_A$ by the rotation through 3 Euler angles 
$\phi _A,\theta _A,-\phi _A$,
as shown in Fig.~\ref{eul-angles}. The angle $\phi _A$ defines the 
orientation 
of the plane containing the $\omega$ direction in the $A$ rest frame 
and the 
$\hat{z_A}$ axis with respect to the 
$\hat{x_A}-\hat{z_A}$ plane. The angle $\theta_A$ is the polar angle of the $\omega$ momentum vector in the $A$ rest frame.
Note that the 
$A$ decay 
plane has an azimuthal angle $\phi _A$ both in the $x_Ay_Az_A$ and 
in the $x_{\omega}y_{\omega}z_{\omega}$ references.
The angles 
$\theta _{\omega}$ and $\phi _{\omega}$ define the orientation of 
the
$\omega$ decay plane in the $\omega$ rest frame.  As the angle 
$\phi _A$ is 
arbitrary, the only angle that has a physical meaning is $\chi = 
\phi _A - \phi _{\omega}$, the opening angle between the $A$ decay 
plane and the $\omega$ decay plane.

Both the $\overline{B}$ meson and the $D$ meson are pseudoscalar, therefore 
their helicity is 0. Thus $A$ will be longitudinally polarized 
independently of its spin. In order to calculate the decay 
amplitude for this $A\to\omega\pi^-$
process, we need to sum over the $\omega$ helicity states:

\begin{equation}
{\cal A} = \Sigma _{\lambda _{\omega}} D^{\star J_A}_{0\lambda
_{\omega}}(\phi _A,\theta _A, -\phi _A)
 D^{\star 1}_{\lambda_{\omega}0}(\phi _{\omega},\theta_{\omega}, -
\phi _
{\omega}) B_{\lambda_{\omega}0},
\end{equation}
here $D^{\star J_A}_{0\lambda_{\omega}}(\phi _A,\theta _A,-\phi _A)$ 
is the 
rotation 
matrix that relates the $x_Ay_Az_A$  and the 
$x_{\omega}y_{\omega}z_{\omega}$ frames
and $ D^{\star 1}_{\lambda_{\omega}0}(\phi 
_{\omega},\theta_{\omega}, -\phi
_{\omega})$ is the rotation matrix relating the 
$x_{\omega}y_{\omega}z_{\omega}$ and the direction of the normal 
to the
$\omega$ decay plane $\hat{n}(\theta _{\omega},\phi_{\omega})$.

In general, there are three helicity amplitudes that contribute to 
this decay:
$B_{10}$ and 
$B_{-10}$, corresponding to a transverse 
$\omega$ polarization, and $B_{00}$, corresponding to a 
longitudinal $\omega$
polarization. This expression can be simplified by observing that 
$A\rightarrow \omega \pi$ is a strong decay and thus conserves
parity. Thus, the helicity amplitudes are related as:

\begin{equation}
B_{10}=(-1)^{1-S(A)}\eta_{A}\eta_{\omega}\eta_{\pi}B_{-10}, 
\label{tran}
\end{equation}
\begin{equation}
B_{00}=(-1)^{1-S(A)}\eta_{A}\eta_{\omega}\eta_{\pi}B_{00}.
\label{long}
\end{equation}
where $S(A)$ is the spin of particle $A$ and $\eta 
_A,\eta_{\omega}$ and
$\eta_{\pi}$ represent the intrinsic parity of the decaying 
particle and its
decay products, respectively.

Eq.~\ref{tran} relates the two transverse helicity
amplitudes, while Eq.~\ref{long} forbids the presence of a 
longitudinal
component under certain conditions. For example, if $A$ is a $1^-$ 
object, the
$\omega$ has transverse polarization and $B_{-10}= - B_{10}$. When 
the sign
in Eqs.~\ref{tran}-\ref{long} is positive, two parameters 
determined by
the hadronic matrix element affect the angular distribution and 
thus we 
cannot fully determine it only on the basis of our assumptions on 
the 
$A$ spin parity. We have carried out the calculation of the 
predicted angular
distributions including spin assignment for $A$ up to 2. The 
predicted
angular distributions are summarized in Table \ref{diff-ang}.

\begin{table}
\caption{Differential angular distributions (modulo a 
proportionality
constant) predicted for different spin assignments. (Note $0^+$ is
forbidden by parity conservation.)}
\label{diff-ang}
\begin{center}
\begin{tabular}{ll}
$J^{P}$ & ${d\sigma}/{d\cos{\theta _A}d\cos{\theta 
_{\omega}}d\chi}$
\\
\hline
$0^-$ &   $|B_{00}|^2 \cos^2{\theta_{\omega}}$  \\
$1^-$ &   $|B_{10}|^2 \sin^2{\theta _A}\sin^2{\theta 
_{\omega}}\sin^2{\chi}$ \\
$1^+$ &   $|B_{10}|^2 \sin^2{\theta _A}\sin^2{\theta 
_{\omega}}^2\cos{\chi}^2 +
           |B_{00}|^2 \cos^2{\theta _A}\cos^2{\theta_{\omega}}$ \\
  &          $~~~~~ - 1/2Re(B_{10}B_{00}^{*})\sin{2\theta _A}
             \sin{2\theta _{\omega}}\cos{\chi}$ \\
$2^-$ & $3|B_{10}|^2\sin^2{2\theta 
_A}\sin^2{\theta_{\omega}}\cos^2{\chi}
        +|B_{00}|^2(3\cos^2{\theta _A}-1)^2\cos^2{\theta 
_{\omega}}$ \\
      & $~~~~~ -\sqrt{3}Re(B_{10}B_{00}^{*})\sin{2\theta 
_A}(3\cos^2{\theta _A}-1)
	\sin{2\theta _{\omega}}\cos{\chi}$\\
$2^+$ &    $3/4 |B_{10}|^2 \sin^2{2\theta _A}\sin^2{\theta 
_{\omega}}\sin^2{\chi}$ 
\\ 
\end{tabular}
\end{center}
\normalsize
\end{table}

The statistical accuracy of our data sample is not sufficient to 
do a 
simultaneous fit of the joint angular distributions shown above. 
Thus only
the projections along the $\theta _A$, $\theta _{\omega}$ and 
$\chi$ are fitted, integrating out the remaining degrees of 
freedom. Table~\ref{proj1},
gives the analytical form for these projections.

\begin{table}
\caption{Projection of the angular distributions along the 
$\cos{\theta _A}$, $\cos{\theta _{\omega}}$ and $\chi$ axes.}
\label{proj1}
\begin{center}
\begin{tabular}{llll}
$J^{P}$ & $d\sigma/d\cos{\theta _A}$ & 
$d\sigma/d\cos{\theta_{\omega}}$
& $d\sigma/d{\chi}$ \\
\hline
$0^-$ & $\frac{4\pi}{3}|B_{00}|^2$  & $4\pi 
|B_{00}|^2\cos^2{\theta _{\omega}}$ & 
$4/3|B_{00}|^2$ \\
$1^-$ & $\frac{4\pi}{3} |B_{10}|^2\sin^2{\theta _A}$&
$\frac{4\pi}{3} |B_{10}|^2\sin^2{\theta_{\omega}}$ &  
$\frac{8}{9}|B_{10}|^2\sin^2{\chi}$\\
$1^+$ & $\frac{4\pi}{3} (|B_{10}|^2 \sin^2{\theta _A}$ 
& $\frac{4\pi}{3} (|B_{10}|^2\sin^2{\theta _{\omega}}$ &  
$ \frac{4}{9}(4|B_{10}|^2\cos^2{\chi}$   \\
  & ~~$+|B_{00}|^2\cos^2{\theta _A})$   &  ~~$+|B_{00}|^2\cos^2{\theta 
_{\omega}})$   &  ~~$+|B_{00}|^2)$   \\
$2^-$ &
$\frac{4\pi}{3}(3|B_{10}|^2\sin^2{2\theta_A}$  
& $\frac{16\pi}{5}(|B_{10}|^2\sin^2{\theta _{\omega}}$
& $4|B_{10}|^2\cos^2{\chi}$    \\
  & ~~$+|B_{00}|^2(3\cos^2{\theta_A}-1)^2)$   & 
~~$+|B_{00}|^2\cos^2{\theta _{\omega}}) $    &  ~~$+|B_{00}|^2$   \\
$2^+$ & $\pi |B_{10}|^2\sin^2{2\theta _A}$ &$\frac{4\pi}{5} 
|B_{10}|^2
\sin^2{\theta _{\omega}}$  & 
$\frac{16}{15}|B_{10}|^2\sin^2{\chi}$ \\
\end{tabular}
\end{center}
\normalsize
\end{table}

\begin{figure}[htb] 
\centerline{\epsfig{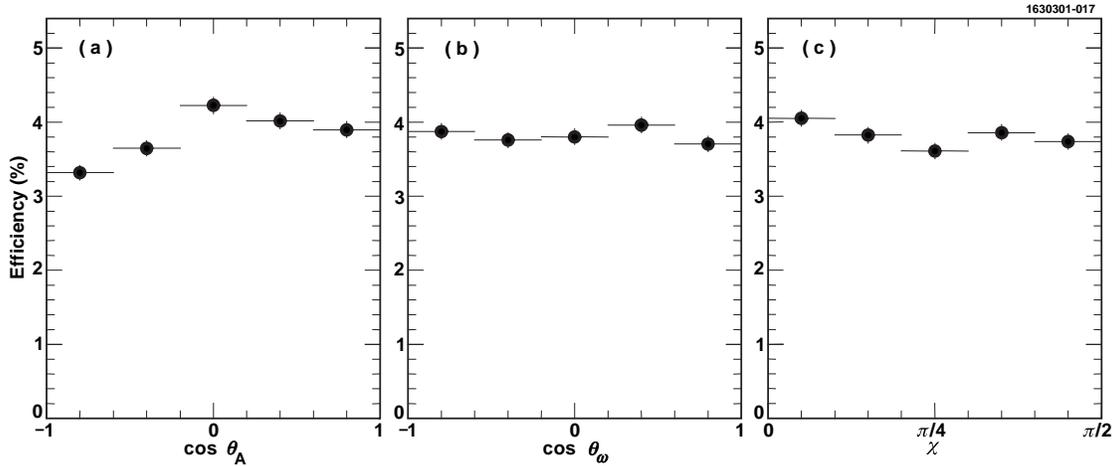}}
\vspace{-1mm} 
\caption{ \label{efficiency} Reconstruction efficiency dependence 
on (a) $\cos\theta_A$, 
(b) $\cos\theta_\omega$, and (c) $\chi$.}
\end{figure}

We determine the projections of these angular distributions by 
fitting  
the $M_B$ distribution as a function of the various angular 
quantities
$\cos\theta_A$, $\cos\theta_{\omega}$, $\chi$. We restrict the 
$\omega\pi^-$
mass range to be between 1.1 and 1.7 GeV, containing 104 signal events. 
In order to fit the angular distribution with theoretical 
expectations,
we must correct the data for acceptances. 
We determine the acceptance correction by comparing the Monte 
Carlo generated
angular distributions with the reconstructed distributions. 
The angular dependent efficiencies are shown in 
Fig.~\ref{efficiency}.
 
\begin{figure}[htbp] 
\centerline{\epsfig{figure=1630301-013.eps,height= 3.2in}\hspace{-.4mm} 
\epsfig{figure=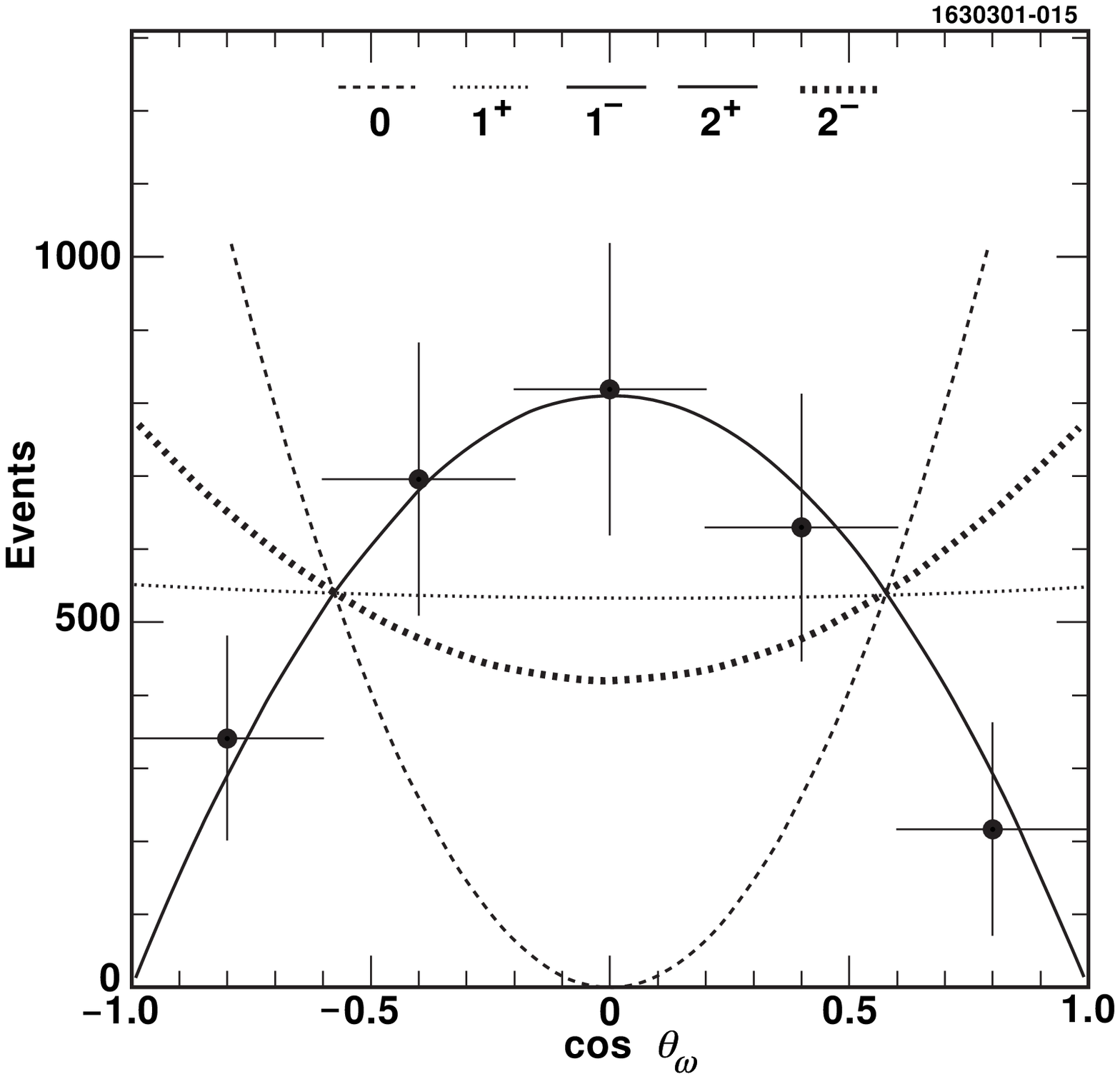,height=3.2in}}
\centerline{\epsfig{figure=1630301-012.eps,height=3.2in}}
\caption{ \label{angle_cosa} The angular distribution of 
$\theta_A$ (top-left),
$\theta_\omega$ (top-right) and $\chi$ (bottom).
The curves show the best fits to the data 
for different $J^P$ assignments. (The $0^-$ and $1^+$ are 
almost indistinguishable in $\cos\theta_A$, while the $1^-$ and 
$2^+$ are indistinguishable in $\cos\theta_{\omega}$ and $\chi$.)}
\end{figure}

The corrected angular distributions are shown in 
Fig.~\ref{angle_cosa}.
The data are fit to the expectations for the various $J^P$ 
assignments. 
For the $0^-$, $1^-$ and $2^+$ assignments, the curves have a 
fixed shape. 
For the $1^+$ and $2^-$ assignments we let the ratio between the 
longitudinal and transverse amplitudes vary to best fit the data. 
We notice that the $\omega$ polarization is very clearly 
transverse 
($\sin^2\theta_{\omega}$) and that infers a $1^-$ or $2^+$ 
assignment.

We list in Table~\ref{table:chi2_of_fit} the $\chi^2/dof$ for the 
different $J^P$ assignments. 
The $1^-$ assignment is preferred, having a  $\chi^2/dof$ of 1.7. 
The other assignments are clearly ruled out. If the $1^-$ assignment
is correct, the probability that our fit yields a $\chi^2/dof$ 
equal to 1.7 or greater is 3.8\% \cite{systematics}.

\begin{table}[hbt]  
\begin{center}  
\caption{Results of fits to angular distributions}
\label{table:chi2_of_fit}  
\begin{tabular}{cccccc}  
              &  $0^-$ &  $1^+$ &  $1^-$ &  $2^+$ & $2^-$ \\\hline  
$\chi^2/dof$ &  7.0 & 4.5 & 1.7 & 3.2 & 5.3 \\
$dof$         &  15 & 14 & 15 & 15 & 14 \\
probability   &  $1.9\times 10^{-15}$   &  $3.3\times 10^{-8}$ & 
3.8\% 
              &  $2.7\times 10^{-5}$    & $3.3\times 10^{-10}$ \\
\end{tabular}  
\end{center}  
\end{table}

\section{Discussion of Nature of the $A^-$}
\label{sec:nature}
We have found a $1^-$ object decaying into $\omega\pi^-$.
A non-relativistic Breit-Wigner fit assuming a single resonance and no 
background gives a mass around 1420 MeV with an intrinsic width 
about 400 MeV. 
Signals for $\omega\pi^-$
resonances have been detected before below 1500 MeV.
There is a well established axial-vector state, the b$_1$(1235), 
with mass 1230 MeV and 
width 142 MeV. Data on vector states, excited $\rho$'s, are 
inconsistent. Clegg and 
Donnachie \cite{Clegg} have reviewed $\tau^-\to(4\pi)^-\bar{\nu}$, 
$e^+e^-\to\pi^+\pi^-$ and $e^+e^-\to\pi^+\pi^+\pi^-\pi^-$ data, 
including the 
$\omega\pi$ final state. Their best explanation is that of two 
$1^-$ states at
1463$\pm$25 MeV and 1730$\pm$30 MeV with widths 311$\pm$62 and 
400$\pm$100 MeV,
respectively. Only the lighter one decays into $\omega\pi$. The 
situation is quite complex, however. They conclude that 
these states must be mixed with non-$q\overline{q}$ states in 
order to explain their decay widths. There is also an observation 
of a wide, 300 MeV, 
$\omega\pi^0$ state in photoproduction at 1250 MeV \cite{Aston}, 
that is dominantly the $b_1$(1235) \cite{Brau} with possibly some
$1^-$ in addition. 
Our state is consistent with the lower mass $\rho'$. We do not
seem to be seeing significant production of the higher mass state 
into $\omega\pi^-$, as expected.

Several models predict the mass and decay widths of excited $\rho$ 
and $\omega$ mesons . For example, according to Godfrey 
and Isgur \cite{GImodel} 
the first radial excitation of the $\rho$ is at 1450 MeV. There is 
a large 
variation among the models, however, on prediction of the relative 
decays widths ranging from no $\pi\pi$ to $\pi\pi$ being equal to 
$\omega\pi$ \cite{Othermodels}.

Since we have observed a wide $1^-$ state in the mass region where 
the $\rho'$ is expected, the most natural explanation is that we 
are observing the $\rho'$ for the first time in $B$ decays.

We note that $\tau^-$ lepton decays into $\omega\pi^-$ have been observed,
and the $1^-$ spin-parity definitely established. 
However, the relatively low mass of the $\tau^-$ distorts the 
the mass spectrum significantly, and makes it difficult to extract the
$\rho'$ mass and width \cite{CLEOtau}\cite{tautopipi}. 

\section{Mass and Width Values for the $\rho'$}
\label{sec:mass}
Here we find the best values for the $\rho'$ mass and width.
This procedure is discussed in more detail in the Appendix.
The shape of the $\omega\pi^-$ mass spectrum is affected by the phase space
allowed by the $B$ decay, the $B$ decay amplitude,
the decay amplitude for the $\rho'$, and finally the shape of the Breit-Wigner decay distribution. For $B\to D^{(*)}\omega\pi^-$ decay:
\begin{eqnarray}
    d\Gamma(B \to D^{(*)}\omega\pi)\ & =&\ \frac{1}{2M_B}\ | A(B\to D^{(*)} \rho^\prime)
    \ BW(\rho^\prime)\ A(\rho^\prime \to \omega \pi)|^2 \\ \nonumber
    & &\times\ d{\cal P}(B\to D^{(*)} \rho^\prime)\ d{\cal P}(\rho^\prime \to \omega \pi)
    \ \frac{dM^2_{\omega\pi}}{2\pi}~~, 
 \label{eq:general0}
 \end{eqnarray}
where $D^{(*)}$ indicates either a $D^*$ or a $D$ meson.

\begin{figure}[htbp]
  \vspace{-0.3in}
   \centerline{\epsfig{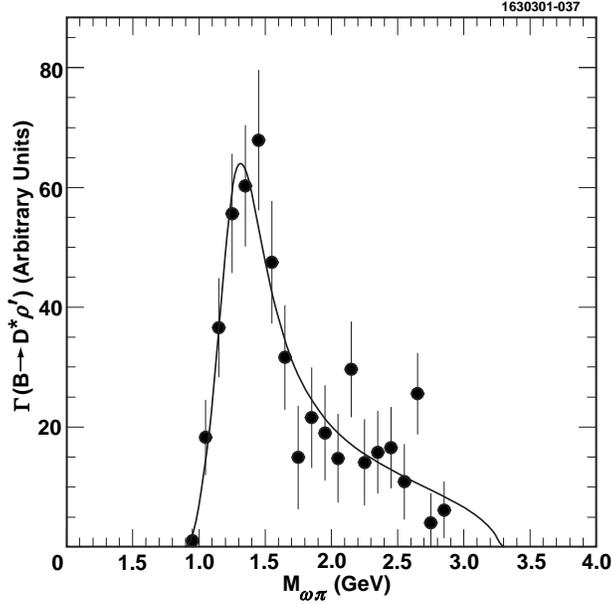}}
  \caption{\label{final_fit} Fit to the $M_{\omega\pi}$ distribution. The
  data in the $D^0$, $D^+$ and $D^{*+}$ channels have been summed and
  corrected using mass dependent efficiencies.}
\end{figure}

The phase space for two-body decays is well known. The decay amplitude for the
$B$ decay can be obtained from factorization where the $\rho'$ is assumed to
be identical to the lepton current \cite{factorization}. Finally the Lorentz structure of the $\rho'$ decay can be accounted for and we are left only to consider a Breit-Wigner amplitude of the form 
\begin{equation}
  {\rm Breit-Wigner}(M_{\omega\pi}) =
  \frac{\Gamma(M_{\omega\pi}) M_{\omega\pi}^2}{(M_{\omega\pi}^2-M_{\rho^\prime}^2)^2
  +M_{\omega\pi}^2\Gamma^2(M_{\omega\pi})}~~,
\end{equation}
where the mass dependent width is given by
\begin{equation}
   \Gamma(M_{\omega\pi}) = \Gamma(M_{\rho^\prime})\ 
\left(\frac{p_\omega(M_{\omega\pi})}{p_\omega(M_{\rho^\prime})}\right)^3
   \ \left(\frac{M_{\rho^\prime}}{M_{\omega\pi}}\right)^n
   \ \frac{1+(Rp_\omega(M_{\rho^\prime}))^2}{1+(Rp_\omega(M_{\omega\pi}))^2}~~.
\end{equation}
We allow the parameters $R$ and $n$ to float in the fit. They represent
parameterizations of the hadronic matrix element and Blatt-Weiskopf damping factors. $R$ is the $\rho'$ radius in units of fm/$\hbar c$.
Fig.~\ref{final_fit} shows the fit to $M_{\omega\pi}$ distribution, where
we have summed the $D^0$, $D^+$ and $D^{*+}$ data. (We have corrected the
data in each channel for the $M_{\omega\pi}$ efficiency dependence, which is
small for the $D^0$ and $D^+$ modes.)

The fit gives the mass and width to be (1349$\pm$25$^{+10}_{-5}$) MeV and
(547$\pm$86$^{+46}_{-45}$) MeV,
respectively. The errors are statistical and systematic; they are derived in
the Appendix. The values for $R$ and $n$ are $2.05^{+7.85}_{-1.14}$ fm/$\hbar c$
and $0.57^{+0.71}_{-0.83}$, respectively.

\begin{figure}[tbh]
\centerline{\epsfig{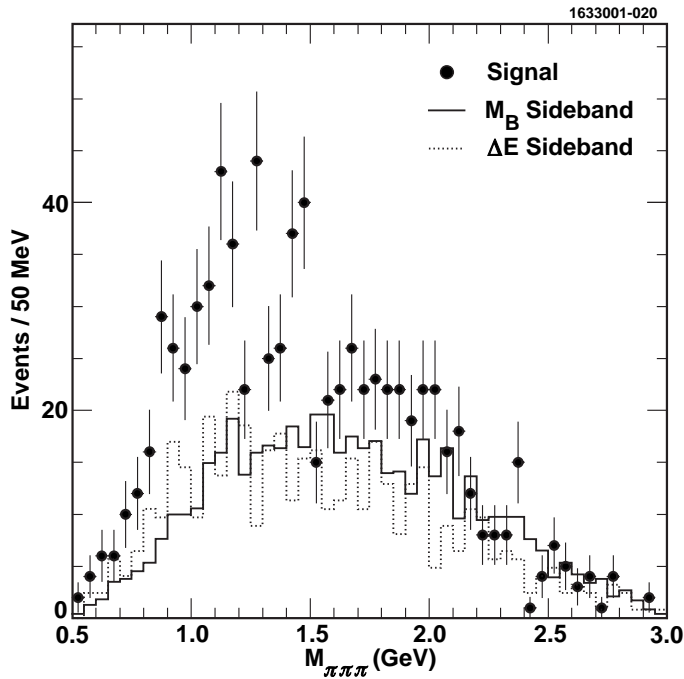}}
\caption{ \label{m3pi_a1_kpi}The invariant mass spectra of 
$\pi^+\pi^-\pi^-$ for the final state
$D^{*+}\pi^+\pi^-\pi^-\pi^0$ for $D^0\to K^-\pi^+$.
The solid histogram is the background
estimate from the $M_B$ lower sideband and the dashed histogram is 
from the
$\Delta E$ sidebands; both are normalized to the fitted number of 
background
events.}
\end{figure}
\section{Search For Other Resonant Substructure in $D^*(4\pi)^-$}
\label{sec:nullsearch}

We have accounted for $\sim$20\% of the $(4\pi)^-$ final state.
We would like to disentangle other resonant substructure.
 Since the background
is large in modes other than $D^0\to K^-\pi^+$ we will only use 
this mode.
One process that comes to mind is that where the virtual $W^-$ 
materializes
as an $a_1^-$, that subsequently decays into $\pi^+\pi^-\pi^-$ and 
we
produce a $D^{**+}$ that decays into $D^{*+}\pi^0$. This process 
should be the
similar to that previously seen in the reaction $B^-\to 
D^{**0}\pi^-$, where
the $D^{**0}$ decayed into a $D^{*+}\pi^-$ \cite{Ddoublestar}.
We search for the presence of an $a_1^-$ by examining
the $\pi^+\pi^-\pi^-$ mass spectrum in
Fig.~\ref{m3pi_a1_kpi}.

\begin{figure}[hbt]
\vspace{-0.5cm}
\centerline{\epsfig{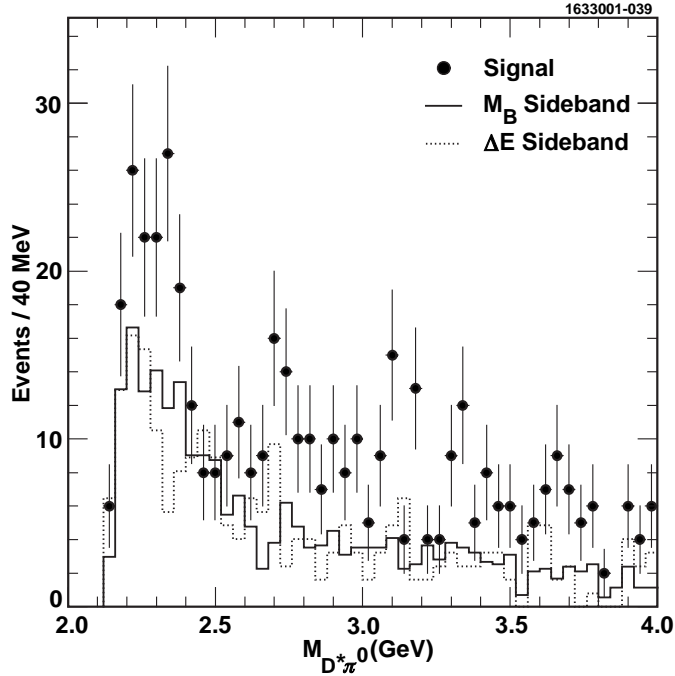}}
\caption{ \label{mdspi0_a1_kpi}The invariant mass spectra of 
$D^{*+}\pi^0$
for $\pi^+\pi^-\pi^-$ masses between 0.6 - 1.6 GeV for the final 
state
$D^{*+}\pi^+\pi^-\pi^-\pi^0$ with $D^0\to K^-\pi^+$.
The solid histogram is the background
estimate from the $M_B$ lower sideband and the dashed histogram is 
from the
$\Delta E$ sidebands; both are normalized to the fitted number of 
background
events.}
\end{figure}

There is an excess of signal events above background in the $a_1^-$ mass
region, that cannot be definitely associated with the $a_1$.
Proceeding by selecting events with $\pi^+\pi^-\pi^-$ masses 
between 0.6
and 1.6 GeV, we show the $D^{*+}\pi^0$ invariant mass spectrum in
Fig.~\ref{mdspi0_a1_kpi}.

Although there is a suggestion of a low mass enhancement, it is 
not consistent
with $D^{**}$ production that would peak in region of 2.42 - 2.46 
GeV. Perhaps
we are seeing an indication of fragmentation at the $b\to c$ decay 
vertex here.

\begin{figure}[htb]
\centerline{\epsfig{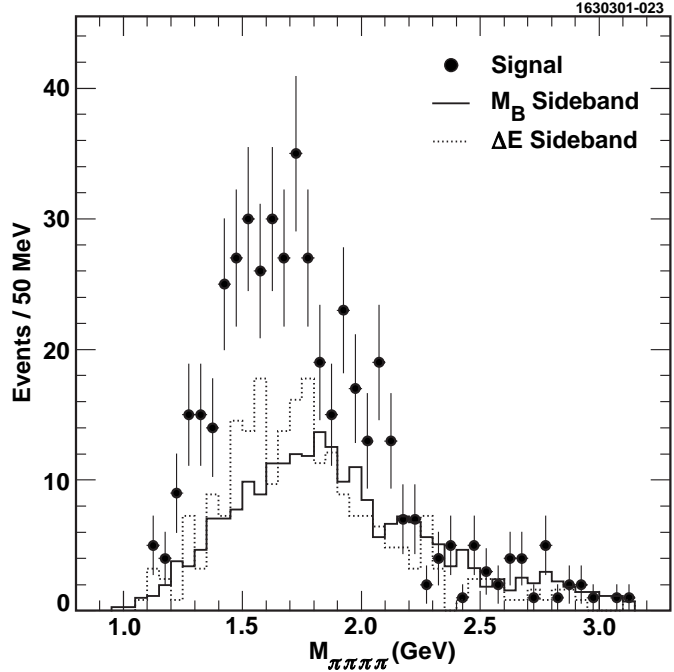}}
\vspace{0.2cm}
\caption{ \label{m4pi_a1_kpi}The invariant mass spectra of
$\pi^+\pi^-\pi^-\pi^0$
for $\pi^+\pi^-\pi^-$ masses between 0.6 - 1.6 GeV for the final 
state
$D^{*+}\pi^+\pi^-\pi^-\pi^0$ with $D^0\to K^-\pi^+$.
The solid histogram is the background
estimate from the $M_B$ lower sideband and the dashed histogram is 
from the
$\Delta E$ sidebands; both are normalized to the fitted number of 
background events.}
\end{figure}

We also display for completeness the ``$a_1^-\pi^0$" mass 
distribution in
Fig.~\ref{m4pi_a1_kpi}.
There may or may not be a wide structure in the $(4\pi)^-$ mass. 
At this point
we abandon our search for substructure in this decay channel.

\section{Discussion and Conclusions}\label{sec:conclusions}

We have made the first statistically significant observations of 
six hadronic $B$ decays shown in Table~\ref{table:brs}. 

\begin{table}[hbt]
\begin{center}
\caption{Measured Branching Ratios}
\label{table:brs}
\begin{tabular}{lcr}
Mode & $\cal{B}$ (\%) & \# of events\\\hline
$\overline{B}^0\to D^{*+}\pi^+\pi^-\pi^-\pi^0$ & 
1.72$\pm$0.14$\pm$0.24   &    1230$\pm$70  \\
$\overline{B}^0\to D^{*+}\omega\pi^-$ & 0.29$\pm$0.03$\pm$0.04& 136$\pm$15  
\\
$\overline{B}^0\to D^{+}\omega\pi^-$ & 0.28$\pm$0.05$\pm$0.04     
&  91$\pm$18  \\
${B}^-\to D^{*0}\pi^+\pi^-\pi^-\pi^0$&1.80$\pm$0.24$\pm$0.27       
&    195$\pm$26   \\
${B}^-\to D^{*0}\omega\pi^-$  & 0.45$\pm$0.10$\pm$0.07       &      
26$\pm$6 \\
${B}^-\to D^{o}\omega\pi^-$  &0.41$\pm$0.07$\pm$0.06    & 
88$\pm$14     \\
\end{tabular}
\end{center}
\end{table}

There is a low-mass resonant substructure in the $\omega\pi^-$ 
mass. The fit to a sophisticated Breit-Wigner function 
gives the mass and width to be (1349$\pm$25$^{+10}_{-5}$) MeV and
(547$\pm$86$^{+46}_{-45}$) MeV,
respectively. 

 The structure at 1349 MeV has a spin-parity 
consistent with $1^-$. It is likely to be the elusive $\rho'$ 
resonance \cite{Clegg}. 
These are by far the most accurate and 
least model dependent measurements of the $\rho'$ parameters.
The $\rho'$ dominates the final state. (Thus the branching ratios 
for the $D^{(*)}\omega\pi^-$ apply also for $D^{(*)}\rho'^-$.)

Heavy quark symmetry predicts equal partial widths for $D^*\rho'$ and $D\rho'$. We measure the relative rates to be
\begin{equation}
{{\Gamma\left(\overline{B}^0\to D^{*+} \rho'^- \right)}\over
{\Gamma\left(\overline{B}^0\to D^{+}\rho'^-\right)}} = 1.04 \pm 
0.21 \pm 0.06 
\end{equation}
\begin{equation}
{{\Gamma\left({B}^-\to D^{*0}\rho'^-\right)}\over
{\Gamma\left({B}^-\to D^0\rho'^-\right)}} = 1.10 \pm 0.31 \pm 
0.06 
\end{equation}
\begin{equation}
{{\Gamma\left(\overline{B}\to D^{*}\rho'^-\right)}\over
{\Gamma\left(\overline{B}\to D\rho'^-\right)}} = 1.06 \pm 0.17 \pm 0.04~~~.
\end{equation}

Thus the prediction of heavy quark symmetry is satisfied within 
our errors. 

Factorization predicts that the fraction of longitudinal  
polarization of the $D^{*+}$ is the same as in the related 
semileptonic decay $\overline{B}\to D^*\ell^-\bar{\nu}$ at four-momentum 
transfer $q^2$ equal to the mass-squared of the $\rho'$ \cite{factorization}
\begin{equation}
{{\Gamma_L\left(\overline{B}\to D^{*+}\rho'^-\right)}\over 
{\Gamma\left(\overline{B}\to D^{*+}\rho'^-\right)}} = 
{{\Gamma_L\left(\overline{B}\to D^{*}\ell^-\bar{\nu}\right)}\over 
{\Gamma\left(\overline{B}\to D^{*}\ell^-
\bar{\nu}\right)}}\left|_{q^2=m^2_{\rho'}}\right.~~.
\end{equation}

Our measurement of the $D^{*+}$ polarization (see Fig.~\ref{cosd}) is 
(63$\pm$9)\%. 
The model predictions in semileptonic decays for a $q^2$ of 2 
GeV$^2$, are between
66.9 and 72.6\% \cite{slmodels}. Thus this prediction of factorization is 
satisfied. 

We can use factorization to estimate the product of the $\rho'$ 
weak decay constant $f_{\rho'}$ 
and the branching ratio for $\rho'^-\to\omega\pi^-$. The relevant 
expression is
\begin{equation}
{{\Gamma\left(\overline{B}\to D^{*+}\rho'^- ,~\rho'^-\to\omega\pi^-\right)}\over 
{{d\Gamma \over dq^2}\left(\overline{B}\to D^*\ell^-\nu\right)|_{q^2=m^2_{\rho'}}}}
=6\pi^2c_1^2f^2_{\rho'}{\cal{B}}\left(\rho'^-\to\omega\pi^-\right)|V_{ud}|^2~~,
\end{equation}
where $c_1$ is a QCD 
correction factor. We use $c_1$=1.1$\pm$0.1 \cite{scaleerr}.

We use the semileptonic decay rates given in Barish \etal ~\cite{Barish}.
The product 
\begin{equation}
f_{\rho'}^2{\cal{B}}\left(\rho'^-\to\omega\pi^-\right)=0.011\pm 0.003 {\rm
~GeV^2}~~,
\end{equation} 
where the error is the quadrature of the experimental errors on the
experimental branching ratios and $c_1$.

The model of Godfrey and Isgur predicts decay constants widths and partial
widths of mesons comprised of light quarks by using a relativistic treatment
in the context of QCD \cite{GImodel}. They predict both $f_{\rho'}$ and
${\cal{B}}\left(\rho'^-\to\omega\pi^-\right)$; the values are 80 MeV and
39\%, respectively. The branching ratio prediction
is believed to be  more accurate \cite{GInotes}. We use this to extract
\begin{equation}
f_{\rho'}=167   \pm 23 {~\rm MeV}~~.
\end{equation}
The model predicts a lower value for $f_{\rho'}$ than observed here, if 
factorization is correct.

We note that all the $B\to D^{(*)}\rho'$ branching ratios that we have 
measured are approximately equal to the $B\to D^{(*)}\rho$ branching rates 
\cite{BigB}
if a model value of ${\cal{B}}\left(\rho'^-\to\omega\pi^-\right)$ = 39\% is
used.

Finally, although the $\overline{B}^0\to D^{*+}(4\pi)^-$ and 
${B}^-\to D^{*0}(4\pi)^-$ branching ratios are nearly equal, the 
$\omega\pi^-$ 
branching ratios are about 1.5 times larger for the charged 
$B$ than the 
neutral $B$, maintaining the trend seen for the $\pi^-$ and 
$\rho^-$ final states. 
Since the $B^-$ lifetime is if anything longer than the $B^0$, 
this trend 
must reverse for some final states. It has not for $D^{(*)}\rho'$.

\section{Acknowledgements}
We thank A. Donnachie, N. Isgur, J. Rosner and J. Schechter for useful discussions.
We also thank D. Black of Syracuse University for crucial
insights on how to calculate Lorentz structures.
We gratefully acknowledge the effort of the CESR staff in providing us with
excellent luminosity and running conditions.
M. Selen thanks the PFF program of the NSF and the Research Corporation, 
and A.H. Mahmood thanks the Texas Advanced Research Program.
This work was supported by the National Science Foundation, the
U.S. Department of Energy, and the Natural Sciences and Engineering Research 
Council of Canada.

\section{Appendix: Determination of the $\rho'$ Mass and Width}
\subsection{Introduction}
Since the $\rho^\prime$ decays to $\omega\pi^-$ via a P-wave,
we need to take into account the fact that the width, $\Gamma(M_{\omega\pi})$,
may not be constant, but can vary with $M_{\omega\pi}$. Furthermore we need to 
consider the kinematic limits from
$B\to D^{(*)}\rho^\prime$ decay and $\rho^\prime \to \omega \pi$ decay.

\subsection {The Differential Decay Distribution}
\vspace*{-0.5pt}
\noindent
We can write a general expression for  
the differential distribution for $M_{\omega\pi}$ in $d\Gamma(B\to D^{(*)}\omega\pi)$
via the $\rho'$ as:
\begin{eqnarray}
    d\Gamma(B \to D^{(*)}\omega\pi)\ =& &\ \frac{1}{2M_B}\ | A(B\to D^{(*)} 
    \rho^\prime)
    \ BW(\rho^\prime)\ A(\rho^\prime \to \omega \pi)|^2 \\ \nonumber
   & \times &\ d{\cal P}(B\to D^{(*)} \rho^\prime)\ d{\cal P}(\rho^\prime \to 
   \omega \pi)
    \ \frac{dM^2_{\omega\pi}}{2\pi}~~,
\label{eq:general}
\end{eqnarray}
 where $d{\cal P}$ indicates a phase space term, $BW$ indicates some form of a Breit-Wigner shape function, and $A$ indicates an amplitude.

The phase space of $B\to D^{(*)} \rho^\prime$ provides a cut off at higher 
$M_{\omega\pi}$,
whereas the phase space of $\rho^\prime \to \omega \pi$ provides a cut off at 
lower $M_{\omega\pi}$.
We assume that the two decay stages are independent and can be factorized.
Thus the function can be calculated from the widths of $B$ and $\rho^\prime$ 
decays.

\subsubsection {The $B$ Decay Width}
\vspace*{-0.5pt}
\noindent
The width for $\rho^\prime$ production is given by:
\begin{eqnarray}
  & &\Gamma(B \to D^{(*)} \rho^\prime)\ =\ \frac{1}{2M_B}\
     |A(B\to D^{(*)} \rho^\prime)|^2\ d{\cal P}(B\to D^* \rho^\prime)\\
  & &d{\cal P}(B\to D^{(*)} \rho^\prime)\ =\ 
  \frac{1}{8\pi}\left(\frac{2p_{D^*}}{M_B}\right)\\\nonumber
  & &A(B\to D^{(*)} \rho^\prime)\ \sim\ \sqrt{G_F} V_{cb} \times
   {\rm Lorentz\ structure} \times f_{\rho'}(M_{\omega\pi})\nonumber
\end{eqnarray}
where $f_{\rho'}(M_{\omega\pi})$ is the $\rho^\prime$ weak production form factor.

Instead of having to calculate $f_{\rho'}(M_{\omega\pi}$), 
we can use our knowledge of semileptonic $b$ decays, $B\to D^{(*)}\ell^-\overline{\nu}$ coupled with factorization to approximate $A(B\to D^* \rho^\prime)$. 

Factorization tells us:
\begin{eqnarray}
    \Gamma(B \to D^{(*)} \rho^\prime)\ &=&\ 6\pi^2\ |V_{ud}|^2\ f^2_{\rho^\prime}(M_{\omega\pi})\
|a_1|^2\ \frac{d\Gamma(B \to D^{(*)} l\nu)}{dq^2}|_{q^2=M_{\omega\pi}^2} \\
\nonumber
     &\propto& p_{D^{(*)}}\ M_{\omega\pi}^2 \times
     \{ |H_+(M_{\omega\pi}^2)|^2 + |H_-(M_{\omega\pi}^2)|^2 + 
     |H_o(M_{\omega\pi}^2)|^2 \}
\label{eq:factorization}
\end{eqnarray}
The helicity amplitudes $H(M_{\omega\pi}^2)$ can be related to the axial-vector form factors
$A_1(q^2)$ and $A_2(q^2)$, and vector form factor $V(q^2)$.
Details can be found in reference~\cite{sheldon}.
\begin{eqnarray}
  H_\pm (q^2)\ &=&\ (M_B+M_{D^*})\ A_1(q^2) \mp \frac{2M_B p_{D^*}}{M_B+M_{D^*}}\ V(q^2)\\\nonumber
  H_o (q^2)\ &=&\ \frac{1}{2M_{D^*}\sqrt{q^2}} \left[ (M_B^2-M_{D^*}^2-q^2)(M_B+M_{D^*})\ A_1(q^2)
  - \frac{4M_B^2 p_{D^*}^2}{M_B+M_{D^*}}\ A_2(q^2) \right]\nonumber
\end{eqnarray}

In the heavy-quark symmetry limit, the form factors $A_1, A_2$, and $V$ are related to the Isgur-Wise function.
With correction due to finite heavy quark mass and $\alpha_s$, they can be
written as
\begin{eqnarray}
  A_1(q^2)\ &=&\ \left[ 1-\frac{q^2}{(M_B+M_{D^*})^2}\right]
              \ \frac{M_B+M_{D^*}}{2\sqrt{M_BM_{D^*}}}\ h_{A_1}(w)\\\nonumber
  A_2(q^2)\ &=&\ R_2\ \frac{M_B+M_{D^*}}{2\sqrt{M_BM_{D^*}}}\
  h_{A_1}(w)\\\nonumber\nonumber
  V(q^2)\   &=&\ R_1\ \frac{M_B+M_{D^*}}{2\sqrt{M_BM_{D^*}}}\
  h_{A_1}(w)~~,\nonumber
\end{eqnarray}
were $w$ is the invariant four-velocity transfer.

The values calculated by Neubert for  $R_1$ and $R_2$ have the explicit dependence on $w$~\cite{neubert} of
\begin{eqnarray}
  R_1(w)\ &=&\ 1.35 - 0.22(w-1) + 0.09(w-1)^2 \\\nonumber
  R_2(w)\ &=&\ 0.79 + 0.15(w-1) - 0.04(w-1)^2~~,
\end{eqnarray}
while Close and Wambach~\cite{close} determine
\begin{eqnarray}
  R_1(w)\ &=&\ 1.15 - 0.07(w-1) + {\cal O} (w-1)^2 \\\nonumber
  R_2(w)\ &=&\ 0.91 + 0.04(w-1) + {\cal O} (w-1)^2~~.
\end{eqnarray}
CLEO measured $R_1(0)$ and $R_2(0)$ to be $1.18\pm 0.30\pm 0.12$
and $0.71\pm 0.22 \pm 0.07$, respectively~\cite{cleodlnu}.
The form factor $h_{A_1}(w)$ can be assumed to be linear as
$h_{A_1}(w) = 1 -\rho_{A_1}^2 (w-1)$; CLEO measured $\rho_{A_1}^2$ to be 
$0.91\pm 0.15\pm 0.06.$

\begin{figure}[htbp]
\vspace{-0.3in}
\centerline{\epsfig{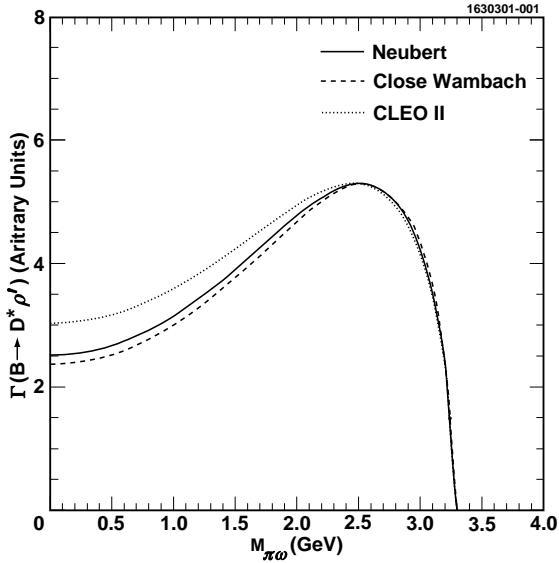}}
\caption{\label{b2dr} Decay width of $B\to D^*\rho^\prime$ as function
      of $M_{\omega\pi}$. The parameters $R_1$ and $R_2$ are from Neubert
      (solid line), Close and Wambach (dashed line) and the CLEO measurement
      (dotted line), with $\rho_{A_1}^2$ as 0.91. }
\end{figure}

$\Gamma(B \to D^* \rho^\prime)$ as function of $M_{\omega\pi}$ is shown in Fig.~\ref{b2dr},
where the form factor $f_{\rho^\prime}$ is not included.

\subsubsection {The Width of $\rho^\prime$ decay}
\vspace*{-0.5pt}
\noindent
The width of $\rho^\prime$ decay can be expanded as:
\begin{eqnarray}
     & &\Gamma(\rho^\prime \to \omega \pi)\ =\ \frac{1}{2M_{\omega\pi} }\
        |A(\rho^\prime \to \omega \pi)|^2\ d{\cal P}(\rho^\prime \to \omega \pi) \\
     & &d{\cal P}(\rho^\prime \to \omega \pi)\ =\ \frac{1}{8\pi}
                                        \left(\frac{2p_\omega}{M_{\omega\pi} }\right)  \\
     & &A(\rho^\prime \to \omega \pi)\ =\ Lorentz\ structure\ \times h(M_{\omega\pi} ^2)~~,
\end{eqnarray}
where $p_\omega$ is evaluated in the $\omega\pi$ rest frame,
and $h(M_{\omega\pi} ^2)$ is the $\rho^\prime$ strong decay form factor,
which is usually assumed to be a constant.

The Lorentz structure of $\rho^\prime \to \omega \pi$ must be linear in
the polarization vectors of the $\rho^\prime$ and the $\omega$.
The simplest mathematical expression is $\varepsilon_{\rho^\prime}\cdot\varepsilon_\omega$,
which is S-wave.
For P-wave, since  $\varepsilon_{\rho^\prime}\cdot p_{\rho^\prime} = 0$ and
$\varepsilon_\omega\cdot p_\omega = 0$ by transversality,
the only Lorentz scalers we can form are
 $(\varepsilon_{\rho^\prime}\cdot p_\omega) (\varepsilon_\omega\cdot p_{\rho^\prime})$
and $\epsilon_{\mu\nu\alpha\beta}\ \varepsilon^\mu_{\rho^\prime}
   \varepsilon^\nu_\omega p^\alpha_\omega p^\beta_{\rho^\prime}$.
The first term violates parity conservation.
Later, we describe the detailed calculation of the second term, which gives
\begin{equation}
    |A(\rho^\prime \to \omega \pi)|^2\ =\ h^2(M_{\omega\pi}^2)
  |\epsilon_{\mu\nu\alpha\beta}\ \varepsilon^\mu_{\rho^\prime}
   \varepsilon^\nu_\omega p^\alpha_\omega p^\beta_{\rho^\prime}|^2
   \ \propto\ h^2(M_{\omega\pi}^2) M_{\omega\pi}^2 p^2_\omega
 \label{eq:lorentz}
\end{equation}

Thus, the width of $\rho^\prime \to \omega \pi$ is:
\begin{equation}
  \Gamma(\rho^\prime \to \omega \pi)\ \propto\ h^2(M_{\omega\pi} ^2) \ p_\omega^3
\end{equation}

Note, that the contribution of longitudinal polarized
$\omega$ to the width is zero, which is consistent with
our spin-parity study; our measurements gave $\Gamma_{Long}/\Gamma_{tot}$ to be
($10\pm 9)\%$ in the $D^*\omega\pi$
mode and ($-0.4\pm22 \%$) in the $D\omega\pi$ mode.

\subsubsection{The Breit-Wigner}
The Breit-Wigner function has a long history. Fundamentally we are
approximating the decay as having a non-changing amplitude as a function of
mass in the simplest case. In general the denominator of the Breit-Wigner
has a fixed form in amplitude given by
\begin{equation}
  {\rm Denominator} = {M_{\omega\pi}^2-M_{\rho^\prime}^2-iM_{\omega\pi}
  \Gamma_{tot}(M_{\omega\pi})}~~,
\end{equation}

The numerator is where we have to have more discussion.
Since $\omega\pi$ is a large decay mode of $\rho^\prime$,
we assume that the mass dependence of $\Gamma_{tot}$ can be
approximated by the  mass dependence of $\Gamma_{\omega\pi}$
\begin{equation}
  \Gamma_{tot}(M_{\omega\pi}) = \Gamma_{tot}(M_{\rho^\prime})\ \frac{\Gamma_{\omega\pi}(M_{\omega\pi})}
  {\Gamma_{\omega\pi}(M_{\rho^\prime})}
\label{eq:bw}
\end{equation}

From the derivation of $\Gamma(\rho^\prime \to \omega \pi)$ above, we see that
\begin{equation}
  \frac{\Gamma_{\omega\pi}(M_{\omega\pi})} {\Gamma_{\omega\pi}(M_{\rho^\prime})} =
  \left(\frac{h(M_{\omega\pi} ^2)}{h(M_{\rho^\prime}^2)}\right)^2 \times
  \left(\frac{p_\omega(M_{\omega\pi})}{p_\omega(M_{\rho^\prime})}\right)^3
\label{eq:run}
\end{equation}

\subsubsection{The Differential Distribution}
Eq.~\ref{eq:general} now can be rewritten as:
\begin{equation}
  d\Gamma(B\to D^*\omega\pi) = \Gamma(B\to D^* \rho^\prime)
  \frac{2M_{\omega\pi} \Gamma_{\omega\pi}(M_{\omega\pi})}{(M_{\omega\pi} ^2-M_{\rho^\prime}^2)^2+M_{\omega\pi}^2\Gamma_{tot}(M_{\omega\pi})^2}
  \frac{dM_{\omega\pi} ^2}{2\pi}
\end{equation}

With the relation shown in Eq.~\ref{eq:bw}, we have the differential function which
can be used for fitting the mass distribution:
\begin{equation}
  \frac{d\Gamma(B\to D^*\omega\pi)}{dM_{\omega\pi}} =C\times \Gamma(B\to D^* \rho^\prime) \times
  \frac{\Gamma M_{\omega\pi}^2}{(M_{\omega\pi}^2-M_{\rho^\prime}^2)^2
  +M_{\omega\pi}^2\Gamma^2}~~,
\end{equation}
where $\Gamma$ is the $M_{\omega\pi}$-dependent total width.
From Eq.~\ref{eq:bw} and Eq.~\ref{eq:run} we get that:
\begin{equation}
     \Gamma = \Gamma_0 \times \left(\frac{h(M_{\omega\pi} ^2)}{h(M_{\rho^\prime}^2)}\right)^2 \times
  \left(\frac{p_\omega(M_{\omega\pi})}{p_\omega(M_{\rho^\prime})}\right)^3
\end{equation}

\subsubsection{Decay Form Factor of $\rho^\prime \to \omega \pi$}
Now, what form should we use for $h(M_{\omega\pi} ^2)$?
In Eq.~\ref{eq:lorentz}, the dimensions of $h(M_{\omega\pi}^2)$ are GeV$^{-1}$.
But the dependence on $M_{\omega\pi} $ could be anything.
So we can try $h(M_{\omega\pi}^2) \propto M_{\omega\pi}^n$, where $n$ is allowed to float in the fit.

$h(M_{\omega\pi}^2)$ could also include Blatt-Weisskopf factors~\cite{blatt}.
A Blatt-Weisskopf factor for the P-wave decay $\rho^\prime \to \omega \pi$ would be of the form:
\begin{equation}
   FF(M_{\omega\pi}^2) = \frac{1+(Rp_\omega(M_{\rho^\prime}))^2}{1+(Rp_\omega(M_{\omega\pi}))^2}
\end{equation}
where $R$ is the radius of the $\rho^\prime$ meson.
A typical value is $R = 1 \ {\rm fm} / \hbar c$.

\subsection{Fit to $M_{\omega\pi}$ Spectrum}

Since we are limited by the statistics, we chose to add the $D$ and $D^*$ 
final states together. This is permissible since
the Lorentz structure of $\rho^\prime$ decay in the different modes is the
same. The only difference is in the width of $B\to D^{(*)}\rho^\prime$ part,
that depends on $M_{\omega\pi}$, however this difference is slight.

We use following formula to represent the width:
\begin{equation}
   \Gamma(M_{\omega\pi}) = \Gamma(M_{\rho^\prime})\ 
\left(\frac{p_\omega(M_{\omega\pi})}{p_\omega(M_{\rho^\prime})}\right)^3
   \ \left(\frac{M_{\rho^\prime}}{M_{\omega\pi}}\right)^n
   \ \frac{1+(Rp_\omega(M_{\rho^\prime}))^2}{1+(Rp_\omega(M_{\omega\pi}))^2}
\end{equation}
We allow the parameters $R$ and $n$ to float. We use Neubert's calculated 
values for $R_1$ and $R_2$; effects of changing this to other estimates will
contribute to the systematic error.
Fig.~\ref{final_fit} shows the fit to $M_{\omega\pi}$ distribution.

The fit gives the mass and width to be (1349$\pm$25) MeV and (547$\pm$86) MeV,
respectively. The values for $R$ and $n$ are $2.05^{+7.85}_{-1.14}$
and $0.57^{+0.71}_{-0.83}$, respectively.

Systematic errors can arise from several sources. We test our fit globally by
restricting the mass range to below 1.8 GeV. This results in a shift in the
mass by -5.7 MeV and the width by -23.2 MeV. Both of these changes are much
smaller than the statistical error.

Two specific sources of systematic error are using the $D^*$
mass to represent both $D^*$ and $D$ meson final states, and
using the form-factors $R_1$ and $R_2$
from Neubert's calculation. To estimate the error on the former, we
use the $D$ in the fit; for the latter 
we use form-factors as calculated by Close and Wambach or,
the CLEO measurement of these parameters.
Listed in the Table~\ref{tb:error} are the contributions from each of these
sources. 

Finally, to estimate the error due to our model of the $\omega\pi^-$
shape, we allow the values of $R$ and $n$ to change so that the $\chi^2$ of
the fit increases by one unit, corresponding to a one standard deviation
variation. We list in Table~\ref{tb:error} the maximum positive and
negative changes in the mass and width values allowed by these variations.
Thus our final value for  the mass of $\rho^\prime$ resonance is
($1349 \pm 25 ^{+10}_{-5}$) MeV.
The width is ($547 \pm 86 ^{+46}_{-45}$) MeV.
\begin{table}[htbp]
   \begin{center}
      \caption{Mass and width fit of resonance $\rho^\prime$}
      \label{tb:error}
      \begin{tabular}{lcc}
         Parameters      & Mass (MeV)    & Width (MeV)  \\ \hline
         Neubert (nominal)& $1349 \pm 25$ & $547 \pm 86$ \\ 
         $ M = M_D$      & $-4.0$        & $-31$        \\ 
         Close-Wambach   & $-1.5$        & $-9$         \\ 
         CLEO $D^*l\nu$  & $+8.2$        & $+24$        \\ \
         Vary $R$ and $n$ & $^{+5.8}_{-3.2}$ & $^{+40}_{-31}$ \\\hline
         Systematic error& $^{+10.0}_{-5.3}$ & $^{+46}_{-45}$ \\ 
      \end{tabular}
   \end{center}
\end{table}


It is possible that values of the $\rho'$ mass and width could be affected if
additional resonance substructure of unknown origin were to be included.
There is no evidence, however, that such structures are needed. By including
all the known physics effects of phase space and Lorentz structure we are able
to describe the data quite well.

\subsection{Calculation of the Lorentz Structure}

In this Section, we will calculate the Lorentz structure
$|\epsilon_{\mu\nu\alpha\beta}\ \varepsilon^\mu_{\rho^\prime}
   \varepsilon^\nu_\omega p^\alpha q^\beta|^2$,
where we use $p^\alpha$ and $q^\beta$ to represent the four momentum
of $\rho^\prime$ and $\omega$.
In general the helicity structure is more apparent when calculated in
a rest frame where the $\rho'$ is in motion.
Later we will give the expression in $\rho^\prime$ rest frame.

The transversality 
$\varepsilon_\omega\cdot p_\omega = 0$ \cite{halzen}.
For spin 1 massive particle with momentum $\vec{p} = p \vec{e}_z$,
the helicity states can be represented as:
\begin{eqnarray}
   \varepsilon_{(\pm)}\ &=&\ \mp(0, 1, \pm i, 0) \\
   \varepsilon_{({\circ})}\ &=&\ (p, 0, 0, E)/M \nonumber
\end{eqnarray}

For general $\vec{p}$, we can construct  $\varepsilon_{({\circ})}$ as follows.
The $\varepsilon_{(\pm)}$ are not shown as they will not be directly used.
\begin{equation}
   \varepsilon_{({\circ})}\ =\ \left( \frac{p}{M},
   \ \frac{E\ \vec{p}}{M\ p} \right)
   \label{eq:hel0}
\end{equation}

Now we need to prepare calculations of several quantities.
First is $\varepsilon_{(\lambda)}^\mu \varepsilon_{(\lambda)}^\nu$.
The summation over all helicity states is given in~\cite{halzen}.
For helicity 0, we use Eq:~\ref{eq:hel0} and expand it.
We have
\begin{eqnarray}
     \sum_{\lambda=o,\pm} \varepsilon_{(\lambda)}^\mu \varepsilon_{(\lambda)}^\nu
        \ &=&\ - g^{\mu\nu} + \frac{p_\mu p_\nu}{M^2} \\
     \varepsilon_{({\circ})}^\mu \varepsilon_{({\circ})}^\nu
        \ &=&\ \frac{p^\mu p^\nu}{M^2}
          - \delta_{\mu 0}\delta_{\nu 0}
          + \frac{(p^\mu p^\nu)^*}{P^2} \\
     \varepsilon_{(+)}^\mu \varepsilon_{(+)}^\nu+
     \varepsilon_{(-)}^\mu \varepsilon_{(-)}^\nu
        \ &=&\ - g^{\mu\nu} + \delta_{\mu 0}\delta_{\nu 0}
          - \frac{(p^\mu p^\nu)^*}{P^2}
\end{eqnarray}
In $(p^\mu p^\nu)^*$, only  $\mu,\nu =1,2,3$ have non-zero values.
In the later calculations, we always assume that transverse helicity states ($\lambda=\pm 1$)
have the same probabilities.

For terms like $\epsilon_{\alpha\beta\gamma\delta}p^\alpha p^\beta\ldots$,
exchanging $\alpha$ and $\beta$ will introduce a negative sign.
Thus the value is zero.
\begin{equation}
    \epsilon_{\alpha\beta\gamma\delta}\ p^\alpha p^\beta\ \times anything\ =\ 0
   \label{eq:four_p}
\end{equation}

The Lorentz structure will be calculated with summation of all helicity states and
different polarizations.
\begin{equation}
A_{tot}  = A_{00} + A_{01} + A_{10} + A_{11}
         = \epsilon_{\alpha\beta\gamma\delta}
         \ \epsilon_{\kappa\lambda\mu\nu}
         \ (\varepsilon^\alpha \varepsilon^\kappa p^\beta p^\lambda)_{\rho^\prime}
         \ (\varepsilon^\gamma \varepsilon^\mu q^\delta p^\nu)_{\omega}
\end{equation}
where $A_{01}$ for longitudinal polarized $\rho^\prime$ and transverse polarized
$\omega$. Others are similar.

There is one term which often shows up; expanding we find:
\begin{eqnarray}
     \epsilon_{0ijk}\epsilon_{0imn}\ p_j p_m \ q_k q_n
     &=&\ p_2^2 q_3^2 + p_3^2 q_2^2 - 2 p_2 p_3 q_2 q_3 \quad (i=1)\\\nonumber
     &+&\ p_1^2 q_3^2 + p_3^2 q_1^2 - 2 p_1 p_3 q_1 q_3 \quad (i=2)\\\nonumber
     &+&\ p_1^2 q_2^2 + p_2^2 q_1^2 - 2 p_1 p_2 q_1 q_2 \quad (i=3)\\\nonumber
     &+&\ ( p_1^2 q_1^2 + p_2^2 q_2^2 + p_3^2 q_3^2 )
     \ -\ ( p_1^2 q_1^2 + p_2^2 q_2^2 + p_3^2 q_3^2 ) \\\nonumber
     &=&\ (p_1^2 + p_2^2 + p_3^2)(q_1^2 + q_2^2 + q_3^2)
    \ -\ (p_1 q_1 + p_2 q_2 + p_3 q_3)^2 \\\nonumber
     & =& p^2 q^2 - (\vec{p}\cdot \vec{q})^2 = p^2 q^2 (1-cos^2\theta)
\end{eqnarray}

Now the summation over all helicity states:
\begin{eqnarray}
 A_{tot}\ &=&\ \epsilon_{\alpha\beta\gamma\delta}
            \ \epsilon_{\kappa\lambda\mu\nu}
            \ \left(-g^{\alpha\kappa}+\frac{p^\alpha p^\kappa}{M^2}\right)
                  p^\beta p^\lambda
            \ \left(-g^{\gamma\mu}+\frac{q^\gamma q^\mu}{m^2}\right)
                 q^\delta q^\nu                      \\\nonumber
          &=&\ \epsilon_{\alpha\beta\gamma\delta}
            \ \epsilon_{\kappa\lambda\mu\nu}
            \ g^{\alpha\kappa} g^{\gamma\mu}
            \ p^\beta p^\lambda q^\delta q^\nu         \\\nonumber
          &=&\ -2 \epsilon_{0 \beta\mu\delta}
            \   \epsilon_{0 \lambda\mu\nu}
            \  p^\beta p^\lambda q^\delta q^\nu     \quad (\alpha=0\ or\
            \kappa=0)\\\nonumber
          & &\quad +\ \epsilon_{\alpha 0 \mu\delta}
            \ \epsilon_{\alpha 0 \mu\nu}
            \ p^0 p^0 q^\delta q^\nu                \quad\ (\beta=\lambda=0)\\\nonumber
          & &\quad +\ \epsilon_{\alpha\beta\nu 0}
            \ \epsilon_{\alpha\lambda\mu 0}
            \  p^\beta p^\lambda q^0 q^0          \quad (\delta=\nu=0)\\\nonumber
          & &\quad +\ 2 \epsilon_{\alpha 0 \mu\delta}
            \   \epsilon_{\alpha \lambda\mu 0}
            \  p^0 p^\lambda q^\delta q^0          \quad (\beta=\nu=0\ or\
            \lambda=\delta=0)\\\nonumber
          &=&\ -2 p^2 q^2 (1-cos^2\theta) + 2 E_{\rho^\prime}^2 q^2
     +  2 E_{\omega}^2 p^2 - 4  E_{\rho^\prime} E_{\omega} p q\ cos\theta
\end{eqnarray}

Lorentz structures for longitudinal polarized $\rho^\prime$ and $\omega$:
\begin{eqnarray}
 A_{00}&=& \epsilon_{\alpha\beta\gamma\delta}
             \epsilon_{\kappa\lambda\mu\nu}
             \left( \frac{p^\alpha p^\kappa}{M^2}
                   - \delta_{\alpha 0}\delta_{\kappa 0}
                   + \frac{(p^\alpha p^\kappa)^*}{p^2}\right)
                  p^\beta p^\lambda
             \left( \frac{q^\gamma q^\mu}{m^2}
                   - \delta_{\gamma 0}\delta_{\mu 0}
                   + \frac{(q^\gamma q^\mu)^*}{q^2}\right)
                 q^\delta q^\nu                      \\\nonumber
          &=& \epsilon_{\alpha\beta\gamma\delta}
             \epsilon_{\kappa\lambda\mu\nu}
                   \ (p^\alpha p^\kappa)^*
                  p^\beta p^\lambda
                   \ (q^\gamma q^\mu)^*
                 q^\delta q^\nu \frac{1}{p^2q^2}
                      \\\nonumber
	  &=& 0
\end{eqnarray}
In the first step, only one of nine terms survive;
seven terms are eliminated due to Eq.~\ref{eq:four_p}.
The one which has four $\delta$'s results in two 0 subscripts in each
$\epsilon$.
In the second step, Eq.~\ref{eq:four_p} limits that
$\beta, \lambda, \delta$ and $\nu$ are all zero,
which also gives zero.

The Lorentz structures for longitudinal polarized $\rho^\prime$ and transverse
polarized $\omega$ are:
\begin{eqnarray}
 A_{01}&=& \epsilon_{\alpha\beta\gamma\delta}
           \epsilon_{\kappa\lambda\mu\nu}
           \left( \frac{p^\alpha p^\kappa}{M^2}
                  - \delta_{\alpha 0}\delta_{\kappa 0}
                  + \frac{(p^\alpha p^\kappa)^*}{p^2}\right)
                  p^\beta p^\lambda
            \left( -g^{\gamma\mu}
                   + \delta_{\gamma 0}\delta_{\mu 0}
                   - \frac{(q^\gamma q^\mu)^*}{q^2}\right)
                 q^\delta q^\nu                      \\\nonumber
          &=& \epsilon_{\alpha\beta\gamma\delta}
            \ \epsilon_{\kappa\lambda\mu\nu}
            \ \delta_{\alpha 0}\delta_{\kappa 0}
            \ g^{\gamma\mu}
            \ p^\beta p^\lambda q^\delta q^\nu
         \ -\ \epsilon_{\alpha\beta\gamma\delta}
            \ \epsilon_{\kappa\lambda\mu\nu}
            \ g^{\gamma\mu}
            \ \frac{(p^\alpha p^\kappa)^*}{p^2}
              p^\beta p^\lambda q^\delta q^\nu           \\\nonumber
          &=& - \epsilon_{0 \beta\mu\delta}
            \ \epsilon_{0 \lambda\mu\nu}
            \ p^\beta p^\lambda q^\delta q^\nu
         \ +\ \epsilon_{\alpha 0 \mu \delta}
            \ \epsilon_{\kappa 0 \mu \nu}
            \ \frac{(p^\alpha p^\kappa)^*}{p^2}
              p^0 p^0 q^\delta q^\nu           \\\nonumber
          &=& (\frac{p^0 p^0}{p^2} -1)
            \ p^2q^2(1-cos^2\theta)             \\\nonumber
          &=& M_{\rho^\prime}^2 q^2 (1-cos^2\theta)~~.
\end{eqnarray}

Similarly, Lorentz structures for transverse polarized $\rho^\prime$ and
longitudinal polarized $\omega$ are:
\begin{equation}
   A_{10}\ =\ m_\omega^2 p^2 (1-cos^2\theta)~~,
\end{equation}
while Lorentz structures for transverse polarized $\rho^\prime$ and $\omega$
are given by:
\begin{eqnarray}
 A_{11}\ &=&\ A_{tot} - A_{00} -A_{01} - A_{10} \nonumber \\
         &=&\ 2 E_{\rho^\prime}^2 q^2 +  2 E_{\omega}^2 p^2
            - 4  E_{\rho^\prime} E_{\omega} p q\ cos\theta
            - (2 p^2 q^2+M_{\rho^\prime}^2 q^2 + m_\omega^2 p^2)
            (1-cos^2\theta) \nonumber \\
         &=&\ (E_{\rho^\prime}^2 + E_{\omega}^2 p^2)\ (1+cos^2\theta)
           \ -\ 4  E_{\rho^\prime} E_{\omega} p q\ cos\theta~~.
\end{eqnarray}

In the rest frame of $\rho^\prime$, the Lorentz structures are expressed as:
\begin{eqnarray}
 A_{tot}\ &=&\ 2M_{\rho^\prime}^2 q_\omega^2   \\\nonumber
 A_{00}\  &=&\ 0   \\\nonumber
 A_{01}\  &=&\ M_{\rho^\prime}^2 q_\omega^2 (1-cos^2\theta)   \\\nonumber
 A_{10}\  &=&\ 0   \\\nonumber
 A_{11}\  &=&\ M_{\rho^\prime}^2 q_\omega^2 (1+cos^2\theta)~~.
\end{eqnarray}
(Note, that there is no contribution from longitudinal polarized
$\omega$.)


\begin{thebibliography}{99}
\bibitem{Differences}
We do not find significant differences between
the CLEO II and II.V data sets.
\bibitem{spineff}
We assume that $\omega\pi$ is in a $1^-$ final state. The 
difference
between the efficiency of a $1^-$ system and an uncorrelated decay
distribution is less than 2\%.
\bibitem{systematics} We allowed a systematic error of 5.5\% on 
the number of
events in each angular bin resulting possibly from the fitting 
procedure
to the $M_B$ spectrum. This changed the probability of $\chi^2$ 
for the 
$1^-$ case to 3.9\% from 3.8\%, and caused negligible changes in the 
other cases.
\bibitem{GInotes}
Private communication from N. Isgur.
The decay constants given in ref.~\cite{GImodel} are dimensionless and must
be multiplied by $m$ to put them in the convention used here.
An additional factor of $\sqrt{2}$ must be used to
convert from the electromagnetic form-factor to the weak form-factor using
CVC.
\end{thebibliography}
\end{document}